\documentclass[3p,twocolumn]{elsarticle}

\usepackage{amssymb}

\usepackage{graphics}

\journal{Astroparticle Physics}

\begin{document}

\begin{frontmatter}



\title{
On the Relation between the True Directions of Neutrinos and
the Reconstructed Directions of Neutrinos in
L/E Analysis Performed by Super-Kamiokande Collaboration Part~1
\\
{\large --- The Effect of Quasi-Elastic Scattering over the Directions of 
the Emitted Leptons with regard to Neutrinos among Fully Contained Events
 ---}
}

\author[HU]{E.~Konishi\corref{cor1}}
\ead{konish@si.hirosaki-u.ac.jp}
\author[KU]{Y.~Minorikawa}
\author[MoU]{V.I.~Galkin}
\author[MeU]{M.~Ishiwata}
\author[SU1]{I.~Nakamura}
\author[HU]{N.~Takahashi}
\author[ky]{M.~Kato}
\author[SU2]{A.~Misaki}
\address[HU]{
Graduate School of Science and Technology, Hirosaki University, Hirosaki, 036-8561, Japan }    
\address[KU]{
Department of Science, School of Science and Engineering, Kinki University, Higashi-Osaka, 577-8502, Japan }
\address[MoU]{
Department of Physics, Moscow State University, Moscow, 119992, Russia}
\address[MeU]{
Department of Physics, Faculty of Science and Technology,
Meisei University, Tokyo, 191-8506, Japan}
\address[SU1]{
Comprehensive Analysis Center for Science, Saitama University, 
Saitama, 338-8570, Japan}
\address[ky]{
Kyowa Interface Science Co.,Ltd., Saitama, 351-0033, Japan }
\address[SU2]{
Inovative Research Organization, Saitama University, Saitama,
 338-8570, Japan}

 \cortext[cor1]{Corresponding author}
\begin{abstract}
It should be regarded that the confirmation of the maximum oscillation in 
neutrino oscillation through $L/E$ analysis by Super-Kamiokande is 
a logical consequence of their establishment on the existence of 
neutrino oscillation through the analysis of the zenith angle 
distribution for atmospheric neutrino events.
 In the present paper (Part~1) with the computer numerical experiment,
 we examine the assumption made by Super-Kamiokande
 Collaboration that the direction of the incident neutrino is
 approximately the same as that of the produced lepton,
which is the cornerstone in their $L/E$ analysis,
 and we find this approximation does not hold even approximately.
In a subsequent paper (Part~2), we apply the results from 
Figures 16, 17, 18 and 19 to $L/E$ analysis and conclude that one cannot
 obtain the maximum oscillation in $L/E$ analysis 
in the single ring muon events due to quasi-elastic scattering
reported by Super-Kamiokande which shows strongly
 the oscillation pattern from the neutrino oscillation.
 
\end{abstract}


\end{frontmatter}


\section{Introduction}

\footnote{ In order to understand the text of our paper well, we strongly 
suggest readers to look at the same paper on the WEB where every figures 
are presented in colors, because figures with colors are strongly 
impressive compare with those with monochrome. In the figures with 
colors, we classify neutrino events by blue points and 
aniti-neutrino events by orange ones.} 
In principle, it is pretty difficult to specify the oscillation 
parameters in neutrino oscillation reliably from the cosmic ray 
experiments (atmospheric neutrinos), even if they really exist, 
because in the cosmic ray experiment by its nature, one cannot 
determine the direction of the incident neutrino which plays a decisive 
role in the analysis of neutrino oscillation and may be determined in the 
accelerator experiments even if the experimental errors are so large due 
to their geometries.\\
\indent On the other hand,
according to the results obtained from the Super-Kamiokande
 Experiments on atmospheric neutrinos, 
it is said that oscillation phenomena have been found between 
two neutrinos, $\nu_{\mu}$ and $\nu_{\tau}$.
 Published reports on the confirmation to the oscillation between the 
neutrinos, $\nu_{\mu}$ and $\nu_{\tau}$, and the history foregoing 
these experiments will be 
critically reviewed and details are in the following:  

\begin{itemize}
\item[(1)]During 1980's Kamiokande and IMB observed the smaller 
atmospheric neutrino flux ratio $\nu_{\mu}/\nu_e$ than the predicted
value \cite{Hirata}.
\item[(2)]Kamiokande found anomaly in the zenith angle distribution \cite{Hatakeyama}.
\item[(3)]Super-Kamiokande found $\nu_{\mu}$-$\nu_{\tau}$ oscillation 
\cite{Kajita2} and 
Soudan2 and MACRO confirmed the Super-Kamiokande result
\cite{Mann}.
\item[(4)]K2K, the first accelerator-based long baseline experiment, 
confirmed atmospheric neutrino oscillation\cite{K2K}.
\item[(5)]MINOS's precision measurement gives the consistent results
with Super-Kamiokande ones\cite{MINOS}.
\end{itemize}

  It is well known 
that Super-Kamiokande Collaboration
examined all possible types of the
neutrino events, such as, say, Sub-GeV e-like, Multi-GeV e-like,
Sub-GeV $\mu$-like, Multi-GeV $\mu$-like, Multi-ring Sub-GeV $\mu$-like,
 Multi-ring 
Multi-GeV $\mu$-like, {\it Upward Stopping Muon Events} and 
{\it Upward Through Going Muon Events}.
In other words, all possible interactions by neutrinos, such as, 
elastic and quasielastic scatterings, single-meson production and
deep scattering are considered in their analyses.
 As the results of them, all topologically 
different types of neutrino events lead to the unified numerical 
oscillation parameters, say,  
$\Delta m^2 = 2.4\times 10^{-3}\rm{ eV^2}$ and $sin^2 2\theta=1.0$
\cite{Ashie2}.\
 
 Really, these parameters are obtained from the unified analysis of the
 zenith angle distributions of various neutrino events.
AS for reliability of the energy estimation which plays a decisively 
important role in the survival probability of a given flavor,
the qualities of the events range in various grades 
from the coarse to the fine.
Generally, it is impossible to estimate the 
individual energy of the muon due to the neutrino interaction among
{\it Partially Contained Events} due to their 
stochastic characters. 
In the multi-ring events,
energy estimations has much uncertainty, 
even if  they belong to {\it Fully Contained Events}. 
Moreover,
it is absolutely impossible to estimate energy of the individual event in 
 {\it Upward Going Muon Events}.


Consequently, in spite of such a fatal defect for the detection of the 
neutrino oscillation, which is inherent in the cosmic ray experiments as 
mentioned above, if one goes ahead and challenges to prove the existence 
of the neutrino oscillation in cosmic ray experiments, then, 
one should focus on analysis of the most reliable events with the 
highest quality exclusively from the experimental point of view, 
discarding all other physical events which have uncertainties more or 
less compared with the events with the highest quality.
 Such the most reliable events among all candidates to be analyzed for 
Super-Kamiokande Collaboration are {\it Fully Contained Events} 
among the single ring muon events due to the quasi elastic scattering 
(QEL). 
These events occupy the majority among {\it Fully Contained events} 
observed by Super-Kamiokande Collaboration.  
By the definition of Fully Contained Events of the single ring events due 
to QEL, 
one can discriminate kinds of neutrinos and can estimate their 
transferred energies because of their confinement in the detector 
as well as their zenith angles which are strongly 
connected with their emitted angles in the neutrino interactions 
concerned. 
If one can find some clear indication on the neutrino oscillation from 
the analysis of the most reliable events with the highest quality, then, 
one can surely expect any corroboration on the neutrino oscillation 
from the analysis of neutrino events with poorer qualities, if 
they really exist. 
This is our motivation for the performance of our computer numerical 
experiment.     

  Here, it should be emphasized strongly that 
Super-Kamiokande Collaboration put a fundamental assumption 
through their whole analyses.
This assumption, however, 
 is never self-evident and, therefore, it should be carefully examined.
 This assumption is that the directions of the incident 
neutrinos are approximately the same as those of emitted leptons,
which we abbreviate as {\it the SK assumption on the direction}
in the present paper. 
This assumption should be recognized as the cornerstone for their
neutrino oscillation analysis throughout all their works,
which links with the survival probability of a given flavor,
$P(\nu_{\mu} \rightarrow \nu_{\mu})$ given by 

\begin{eqnarray}
\lefteqn{P(\nu_{\mu} \rightarrow \nu_{\mu})=}
 \nonumber \\
&& 1- sin^2 2\theta \cdot sin^2
(1.27\Delta m^2 L_{\nu} / E_{\nu} ),  
\\ \nonumber 
\end{eqnarray}                                                    

where $sin^2 2\theta = 1.0$ and
$\Delta m^2 = 2.4\times 10^{-3}\rm{eV^2}$\cite{Ashie2}.

The most important result among the achievements on neutrino oscillation
 made by Super-Kamiokande Collaboration is the finding of the 
maximum oscillation 
in neutrino oscillation, because it is the ultimate result
in the sense that they observe the oscillation pattern itself
directly in neutrino oscillation.

Therefore, it is desirable to carry out careful $L/E$ analysis for 
the single ring muon events due to QEL 
(the events with the highest quality) among 
{\it Fully Contained Events}  exclusively in order to obtain 
a definite conclusion around the neutrino oscillation,
 focusing on the validity (or invalidity) of 
{\it the SK assumption on the direction}.
  Consequently, let us examine 
{\it the SK assumption on the direction}.
 This is the main theme of the present paper (Part~1).
 In a subsequent paper (Part~2), we discuss the application of the 
result obtained in Part~1 to the analysis for the single ring muon events 
due to QEL among {\it Fully Contained Events},
concluding the invalidity of
{\it the SK assumption on the direction}
through $L/E$ analysis.

In order to avoid any misunderstanding toward 
{\it the SK assumption on the direction}, we reproduce this assumption 
from the original SK papers and their related papers in italic.

[A] Kajita and Totsuka \cite{Kajita1} state
\footnote{see page 101 of the paper concerned.}:
\begin{quote}
"{\it However, the direction of the neutrino must be estimated from the
reconstructed direction of the products of the neutrino interaction.
 In water Cheren-kov detectors, the direction of an observed lepton is
assumed to be the direction of the neutrino. Fig.11 and 
Fig.12 show the estimated correlation angle between 
neutrinos and leptons as a function of lepton momentum.
 At energies below 400~MeV/c, the lepton direction has little 
correlation with the neutrino direction. The correlation angle 
becomes smaller with increasing lepton momentum. Therefore, 
the zenith angle dependence of the flux as a consequence of 
neutrino oscillation is largely washed out below~400 MeV/c lepton
momentum. With increasing momentum, the effect can be seen more 
clearly.}" 
\end{quote}

[B] Ishitsuka \cite{Ishitsuka} states\footnote
{see page 138 of the paper concerned.}:
\begin{quote}
" {\it 8.4  Reconstruction of $L_\nu$
\vskip 2mm

Flight length of neutrino is determined from the neutrino incident
zenith angle, although the energy and the flavor are also involved.
 First, the direction of neutrino is estimated for each sample by 
a different way. Then, the neutrino flight lenght is 
calclulated from the zenith angle of the reconstructed direction.
\\
\\
 8.4.1 Reconstruction of Neutrino Direction

\vspace{-2mm}

{\flushleft{\underline 
{FC Single-ring Sample}
}
}

\vspace{2mm}

The direction of neutrino for FC single-ring sample is 
simply assumed to be the same as the reconstructed direction of muon.
Zenith angle of neutrino is reconstructed as follows:
\[
\hspace{0.5cm}\cos\Theta^{rec}_{\nu}=\cos\Theta_{\mu} \hspace{1cm}(8.17) 
\]
,where $\cos\Theta^{rec}_{\nu}$ and $\cos\Theta_{\mu}$ are 
cosine of the reconstructed zenith angle of neutrino and muon,
respectively.}" 
\end{quote}

[C] Jung, Kajita {\it et al.} \cite{Jung} state
\footnote{see page 453 of the paper concerned.}:
\begin{quote}
"{\it At neutrino energies of more than a few hundred MeV, the 
direction of the reconstructed lepton approximately represents 
the direction of the original neutrino. Hence, for detectors near 
the surface of the Earth, the neutrino flight distance is a function of
the zenith direction of the lepton. Any effects, 
such as neutrino oscillations, that are a function of the neutrino
 flight distance will be manifest in the 
lepton zenith angle distributions.}" 
\end{quote}

As clarified from Figures~11, 12 and 26 in \cite{Kajita1} and Figure 8.12 
in \cite{Ishitsuka}, 
Super-Kamiokande Collaboration know well the  existence of non-negligible 
scattering angles in the neutrino interactions.
Nevertheless, the reason why they put the fundamental postulate on the 
direction mentioned above is surmised to be due to their recognition that 
statistically enough accumulation of the events leads to the 
$\cos\theta_{\nu}=\cos\theta_{\mu}$ 
as a whole after mutual cancellation of the scattered angles. 

As already stated, among all possible events to be analyzed
 the most important events from which we can extract 
neutrino oscillation parameters definitly
are undoubtedly the single ring muon (electron) events
 which are generated in the detector and terminate in it, 
because these events give the essential information for clear
 interpretation on neutrino oscillation, if it really exists,
 namely, the kinds of the
neutrinos, the transferred energies to the charged particles and 
their directions. 
The single muon events are generated mainly from the quasi-elastic 
scattering(QEL). 
If the existence of neutrino oscillation is verified definitely 
in the analysis of single ring muon events among
{\it Fully Contained Events} under 
{\it the SK assumption on the direction}, 
then again we can say, we expect 
the corroborations even in the analysis of other types of the 
interactions with poorer qualities.
 Therefore, the analysis for the single ring muon events among 
{\it Fully Contained Events} is decisively important compared with 
the analysis of other types of the neutrino events. 
Consequently, first of all, let us start to examine the validity on 
{\it the SK assumption on the direction}.    

Our paper is organized as follows.
 In section~2, we treat the differential cross section for QEL in the
stochastic manner as exactly as possible and obtain the zenith angle 
distributions of the emitted leptons, $\cos\theta_\mu$,
 for given neutrinos with definite zenith angles, taking account of 
the azimuthal angles of the emitted leptons in QEL.
 As a result of it, we show that 
{\it the SK assumption on the direction} does not hold any more for 
the incident neutrinos with smaller energies. 

In section~3,  we give the correlation between 
$cos\theta_{\nu(\bar{\nu})}$ and $cos\theta_{\mu(\bar{\mu})}$ or 
the correlation between $L_{\nu(\bar{\nu})}$ and $L_{\mu(\bar{\mu})}$ 
and mention the reason why {\it the SK assumption on the direction} 
does not hold. 

\section{Single Ring Events among Fully Contained 
Events which are Produced by Quasi Elastic Scsattering.
}

\subsection{
Differential cross section of quasi elastic scattering and 
spreads of the scattering angles
}
 As stated in Introduction, the finding of observation of 
the maximum oscillation in the $L/E$ analysis is the ultimate 
verification of the finding of neutrino 
oscillation by Super-Kamiokande. 
 For the examination of the Super-Kamiokande's assertion, we analyze 
 $L/E$ distribution of the single ring events among 
{\it Fully Contained Events}.

 In order to examine the validity of {\it the SK assumption on 
the direction},  we consider the single ring events due to
the following quasi elastic scattering(QEL):

   \begin{eqnarray}
         \nu_e + n \longrightarrow p + e^-  \nonumber\\
         \nu_{\mu} + n \longrightarrow p + \mu^- \nonumber\\
         \bar{\nu}_e + p \longrightarrow n + e^+ \\
         \bar{\nu}_{\mu}+ p \longrightarrow n + \mu^+ \nonumber
         ,\label{eqn:1}
   \end{eqnarray}
The differential cross section for QEL is given as follows \cite{r4}.\\
    \begin{eqnarray}
         \frac{ {\rm d}\sigma_{\ell(\bar{\ell})}(E_{\nu(\bar{\nu})}) }{{\rm d}Q^2} = \nonumber\\
         \frac{G_F^2{\rm cos}^2 \theta_C}{8\pi E_{\nu(\bar{\nu})}^2}
         \Biggl\{ A(Q^2) \pm B(Q^2) \biggl[ \frac{s-u}{M^2} \biggr]
          \nonumber \\ 
+ C(Q^2) \biggl[ \frac{s-u}{M^2} \biggr]^2 \Biggr\},
         \label{eqn:2}
    \end{eqnarray}

\noindent where
    \begin{eqnarray*}
      A(Q^2) &=& \frac{Q^2}{4}\Biggl[ f^2_1\biggl( \frac{Q^2}{M^2}-4 \biggr)+ f_1f_2 \frac{4Q^2}{M^2} \\
 &&+  f_2^2\biggl( \frac{Q^2}{M^2} -\frac{Q^4}{4M^4} \biggr) + g_1^2\biggl( 4+\frac{Q^2}{M^2} \biggl) \Biggr], \\
      B(Q^2) &=& (f_1+f_2)g_1Q^2, \\
      C(Q^2) &=& \frac{M^2}{4}\biggl( f^2_1+f^2_2\frac{Q^2}{4M^2}+g_1^2 \biggr).
    \end{eqnarray*}
\\
\noindent The signs $+$ and $-$ refer to $\nu_{\mu(e)}$ and $\bar{\nu}_{\mu(e)}$ for charged current (c.c.) interactions, respectively.  The $Q^2$ denotes the four momentum transfer between the incident neutrino and the charged lepton. Details of other symbols are given in \cite{r4}.

The relation among $Q^2$, $E_{\nu(\bar{\nu})}$, energy of the incident 
neutrino, $E_{\ell}$, energy of the emitted charged lepton (muon or 
electron or their anti-particles) and $\theta_{\rm s}$, scattering angle 
of the emitted lepton, is given as
      \begin{equation}
         Q^2 = 2E_{\nu(\bar{\nu})}E_{\ell}(1-{\rm cos}\theta_{\rm s}).
                  \label{eqn:3}
      \end{equation}

\noindent Also, energy of the emitted lepton is given by
      \begin{equation}
         E_{\ell} = E_{\nu(\bar{\nu})} - \frac{Q^2}{2M}.
\label{eqn:4}
      \end{equation}

\begin{figure}
\begin{center}
\vspace{-1.0cm}
\hspace*{-0.8cm}
\rotatebox{90}{%
\resizebox{0.35\textwidth}{!}{\includegraphics{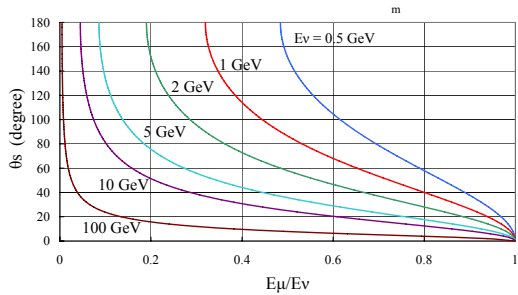}}}

\vspace{-1cm}
\caption{Relation between the energy of the muon and its 
scattering angle for different incident muon neutrino energies,
 0.5, 1, 2, 5, 10 and 100~GeV.}
\label{figH001}
\end{center}
\end{figure} 

Now, let us examine magnitude of the scattering angle of the emitted 
lepton in a quantitative way, as this plays a decisive role in determining 
the accuracy of direction of the incident neutrino,
which is directly related to reliability of the zenith angle 
distribution of 
single ring muon (electron) events in the Super-Kamiokande Experiment.
By using Eqs. (\ref{eqn:2}) to (\ref{eqn:4}), we obtain the distribution 
function for scattering angle of the emitted leptons and the related 
quantities by a Monte Carlo method. The procedure for determining 
scattering angle for a given energy of the incident neutrino is described 
in Appendix A.  Figure~\ref{figH001} shows this relation for muon, from 
which we can easily understand that the scattering angle $\theta_{\rm s}$ 
of the emitted lepton (muon here) cannot be neglected.  For a 
quantitative examination of the scattering angle, we construct the 
distribution function for ${\theta_{\rm s}}$ of the emitted lepton from 
Eqs. (\ref{eqn:2}) to (\ref{eqn:4}) by using the Monte Carlo method.

\begin{figure}
\begin{center}
\vspace{-3.0cm}
\hspace*{-0.8cm}
\rotatebox{90}{%
\resizebox{0.45\textwidth}{!}{\includegraphics{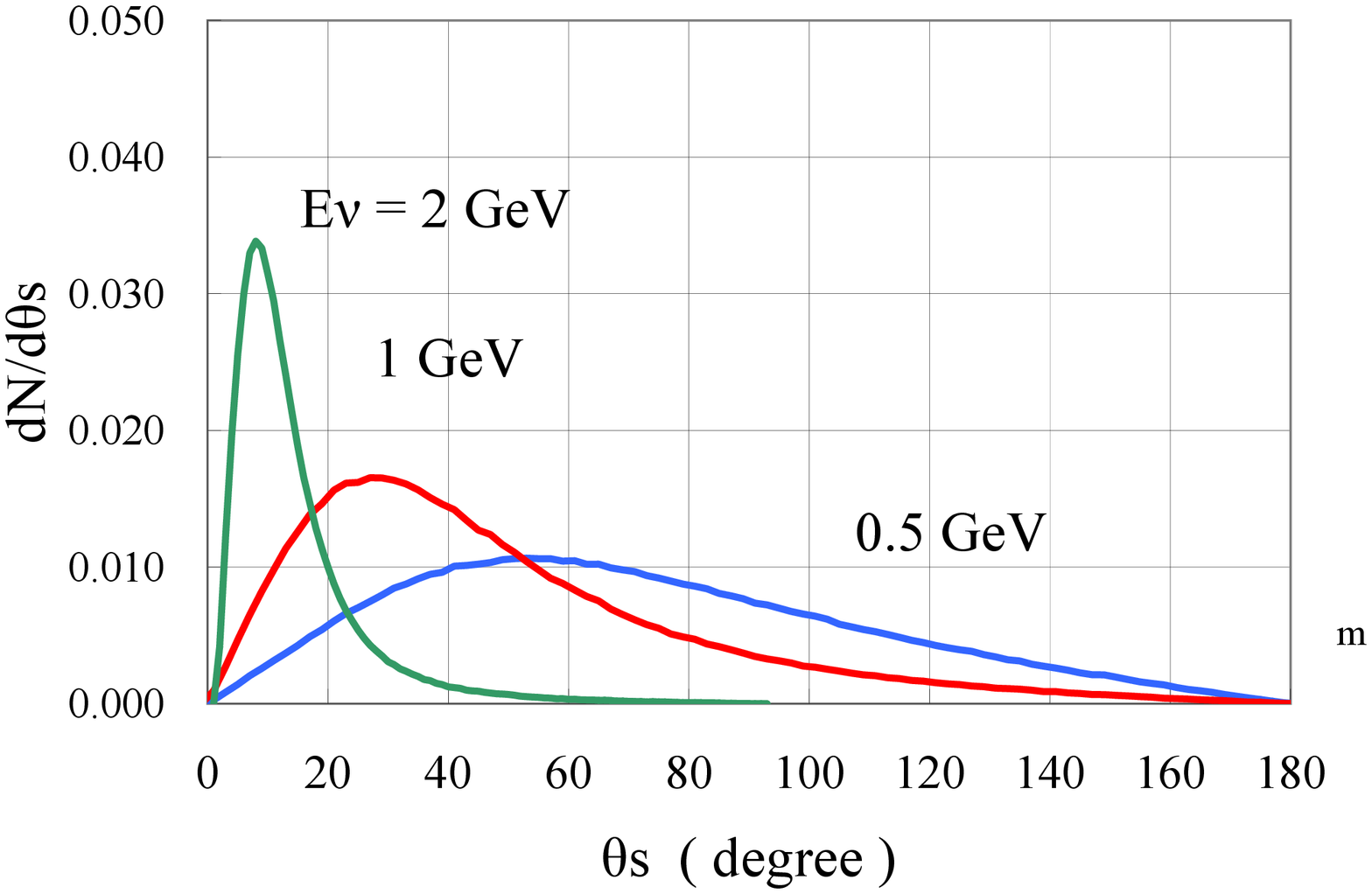}}}
\vspace{-1.5cm}
\caption{Distribution functions for the scattering angle of 
the muon for muon-neutrino with incident energies, 0.5 , 1.0 and 
2~GeV. Each curve is obtained by the Monte Carlo 
method (one million sampling per each curve). }
\label{figH002}

\resizebox{0.5\textwidth}{!}{%
  \includegraphics{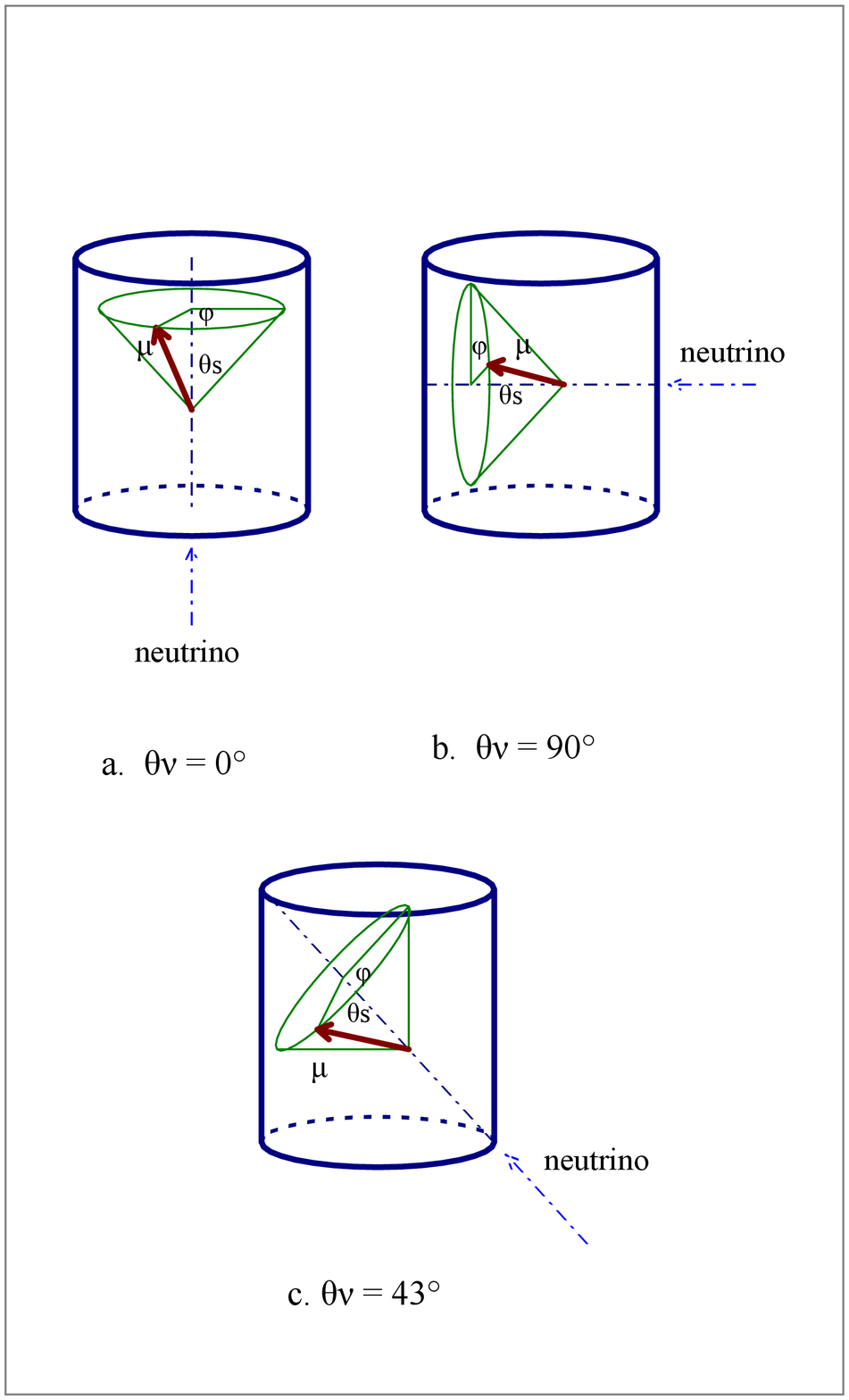}
  }
\end{center}
\vspace{-1.0cm}
\caption{Schematic view of the zenith angles of the charged
 muons for different zenith angles of the incident neutrinos, focusing on
  their azimuthal angles.}
\label{figH003}
\end{figure}

Figure~\ref{figH002} gives the distribution function for $\theta_{\rm s}$ 
of the muons produced in the muon neutrino interactions.
 It can be seen that the muons produced from lower energy neutrinos are 
scattered over wider angles and that a considerable part of them are
 scattered even in backward directions. 
Similar results are obtained for anti-muon neutrinos, electron neutrinos 
and anti-electron neutrinos.

%

Also, in a similar manner, we obtain not only the distribution function 
for the scattering angle of the charged leptons, but also their average 
values $<\theta_{\rm s}>$ and their standard deviations $\sigma_{\rm s}$. 
Table~1 shows them  for muon neutrinos, anti-muon neutrinos, electron 
neutrinos and anti-electron neutrinos.  
From Table~1, it seems to be clear that the scattering angles 
could not be neglected, taking account of the fact that
 the frequency of the neutrino events with smaller energies is far 
larger than that of the neutrino events with larger energies 
due to high steep of the neutrino energy spectrum.  
 However,   
Super-Kamiokande Collaboration assume that the direction of the neutrino 
is approximately the same as that of the emitted lepton even for the 
neutrino events with smaller energies,
as cited in {\it the three
passages} mentioned above \cite{Kajita1},
\cite{Ishitsuka},\cite{Jung}. 
However, this assumption has never been verified by Super-Kamiokande 
Collaboration.

\begin{table*}
\vspace{-2mm}
\caption{\label{tab:table1} The average values $<\theta_{\rm s}>$ for 
scattering angle of the emitted charged leptons and their standard 
deviations $\sigma_{\rm s}$ for various primary neutrino energies 
$E_{\nu(\bar{\nu})}$.}
\vspace{5mm}
\begin{center}
\begin{tabular}{|c|c|c|c|c|c|}
\hline
&&&&&\\
$E_{\nu(\bar{\nu})}$ (GeV)&angle&$\nu_{\mu(\bar{\mu})}$&
$\bar{\nu}_{\mu(\bar{\mu})}$&$\nu_e$&$\bar{\nu_e}$ \\
&(degree)&&&&\\
\hline
0.2&$<\theta_\mathrm{s}>$&~~ 89.86 ~~&~~ 67.29 ~~&~~ 89.74 ~~&~~ 67.47 ~~\\
\cline{2-6}
   & $\sigma_\mathrm{s}$ & 38.63 & 36.39 & 38.65 & 36.45 \\
\hline
0.5&$<\theta_\mathrm{s}>$& 72.17 & 50.71 & 72.12 & 50.78 \\
\cline{2-6}
   & $\sigma_\mathrm{s}$ & 37.08 & 32.79 & 37.08 & 32.82 \\
\hline
1  &$<\theta_\mathrm{s}>$& 48.44 & 36.00 & 48.42 & 36.01 \\
\cline{2-6}
   & $\sigma_\mathrm{s}$ & 32.07 & 27.05 & 32.06 & 27.05 \\
\hline
2  &$<\theta_\mathrm{s}>$& 25.84 & 20.20 & 25.84 & 20.20 \\
\cline{2-6}
   & $\sigma_\mathrm{s}$ & 21.40 & 17.04 & 21.40 & 17.04 \\
\hline
5  &$<\theta_\mathrm{s}>$&  8.84 &  7.87 &  8.84 &  7.87 \\
\cline{2-6}
   & $\sigma_\mathrm{s}$ &  8.01 &  7.33 &  8.01 &  7.33 \\
\hline
10 &$<\theta_\mathrm{s}>$&  4.14 &  3.82 &  4.14 &  3.82 \\
\cline{2-6}
   & $\sigma_\mathrm{s}$ &  3.71 &  3.22 &  3.71 &  3.22 \\
\hline
100&$<\theta_\mathrm{s}>$&  0.38 &  0.39 &  0.38 &  0.39 \\
\cline{2-6}
   & $\sigma_\mathrm{s}$ &  0.23 &  0.24 &  0.23 &  0.24 \\
\hline
\end{tabular}
\end{center}
\label{tab:1}
\end{table*}

\subsection{Influence of the azimuthal angle in QEL
 over the zenith angle of single ring events}

In the present subsection, we examine the effect of the azimuthal angles,
$\phi$, of emitted leptons over their own zenith angles,
$\theta_{\mu(\bar{\mu}))}$, for given zenith angles 
of the incident neutrinos, $\theta_{\nu(\bar{\nu}))}$ in QEL,
 which was not be considered in the detector simulation
carried by Super-Kamiokande Collaboration
\footnote{Throughout this paper, we measure the zenith angles of the 
emitted leptons from the upward vertical direction of the incident 
neutrino. Consequently, notice that the sign of our direction is opposite 
to that of the Super-Kamiokande Experiment 
( our $\cos\theta_{\nu(\bar{\nu})}$ = - 
$\cos\theta_{\nu(\bar{\nu})}$ in SK)}.
The influence of this effect over the zenith angle cannot be neglected
particularly in horizontal-like neutrino events.

For three typical cases (vertical, horizontal and diagonal), 
Figure~\ref{figH003} gives 
a schematic representation of the relationship between, 
$\theta_{\nu(\bar{\nu})}$, the zenith angle of the incident neutrino, 
and ($\theta_{\rm s}$, $\phi$),
 a pair of scattering angle of the emitted muon and its azimuthal angle.
Zenith angle of the emitted muon is derived from 
$\theta_{\nu(\bar{\nu})}$ and ($\theta_{\rm s}$, $\phi$) by (A.6)
as shown in Appendix A.  

From Figure~\ref{figH003}-a, it can been seen that the zenith angle 
$\theta_{\mu(\bar{\mu})}$ of the emitted lepton is not influenced by its 
$\phi$ in the vertical incidence of the neutrinos 
$(\theta_{\nu(\bar{\nu})}=0^{\rm o})$, as it must be. From 
Figure~\ref{figH003}-b, 
however, it is obvious that the influence of $\phi$ of the emitted leptons 
on their own zenith angle is the strongest in the case of horizontal 
incidence of the neutrino $(\theta_{\nu(\bar{\nu})}=90^{\rm o})$. Namely, 
one half of the emitted leptons are recognized as upward going, while the 
other half is classified as downward going ones. The diagonal case ( 
$\theta_{\nu(\bar{\nu})}=43^{\rm o}$) is intermediate between the vertical 
and the horizontal. In the following, we examine the cases for vertical, 
horizontal and diagonal incidence of the neutrinos with different
energies, 
say, $E_{\nu(\bar{\nu})}=0.5$ GeV, $E_{\nu(\bar{\nu})}=1$ GeV and 
$E_{\nu(\bar{\nu})}=5$ GeV,
as the typical cases.

\begin{figure*}
\vspace{-1.0cm}
\hspace{2.5cm}(a)
\hspace{5.5cm}(b)
\hspace{5.5cm}(c)
\vspace{-0.3cm}
\begin{center}
\resizebox{\textwidth}{!}{%
  \includegraphics{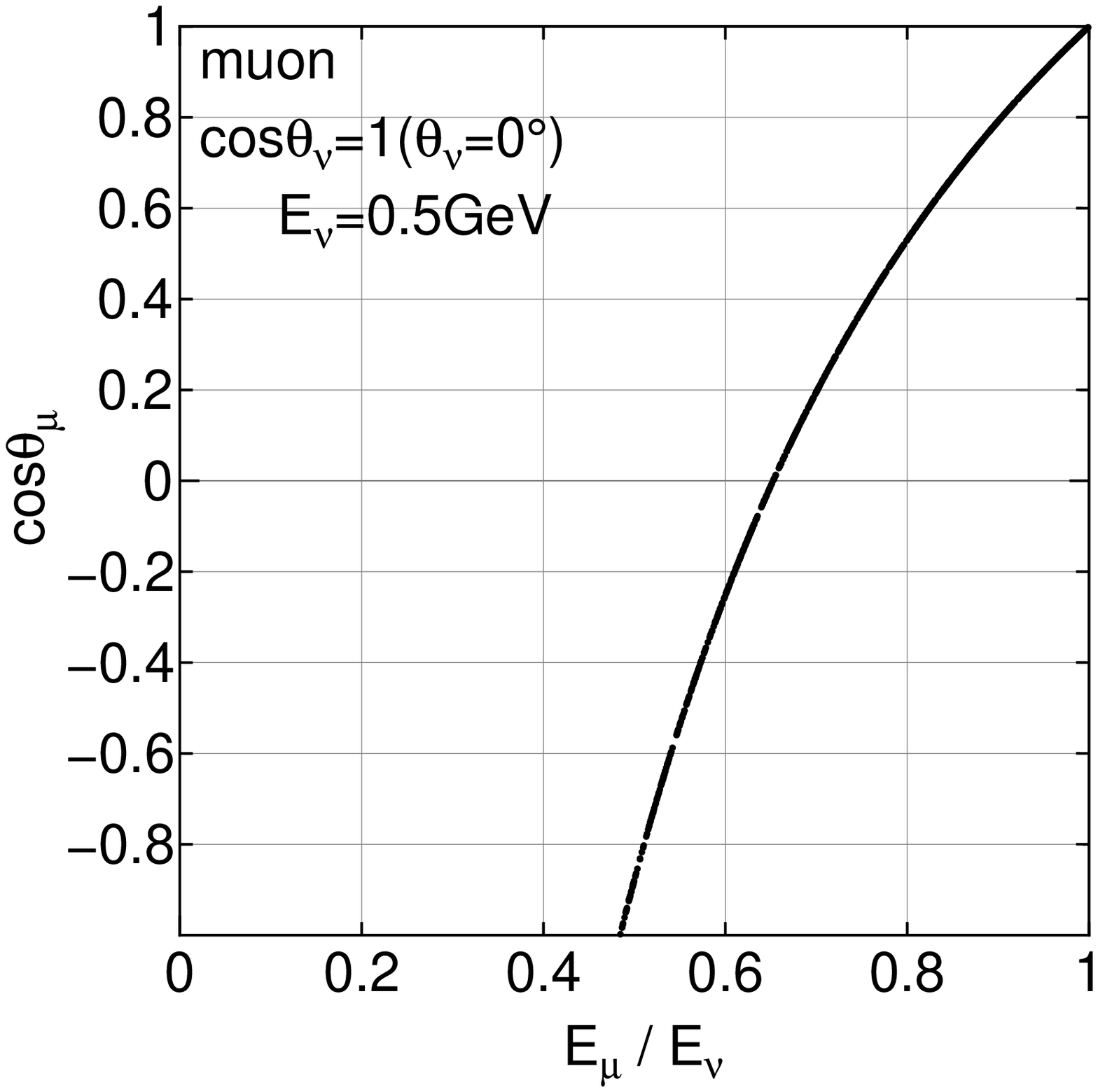}\hspace{1cm}
  \includegraphics{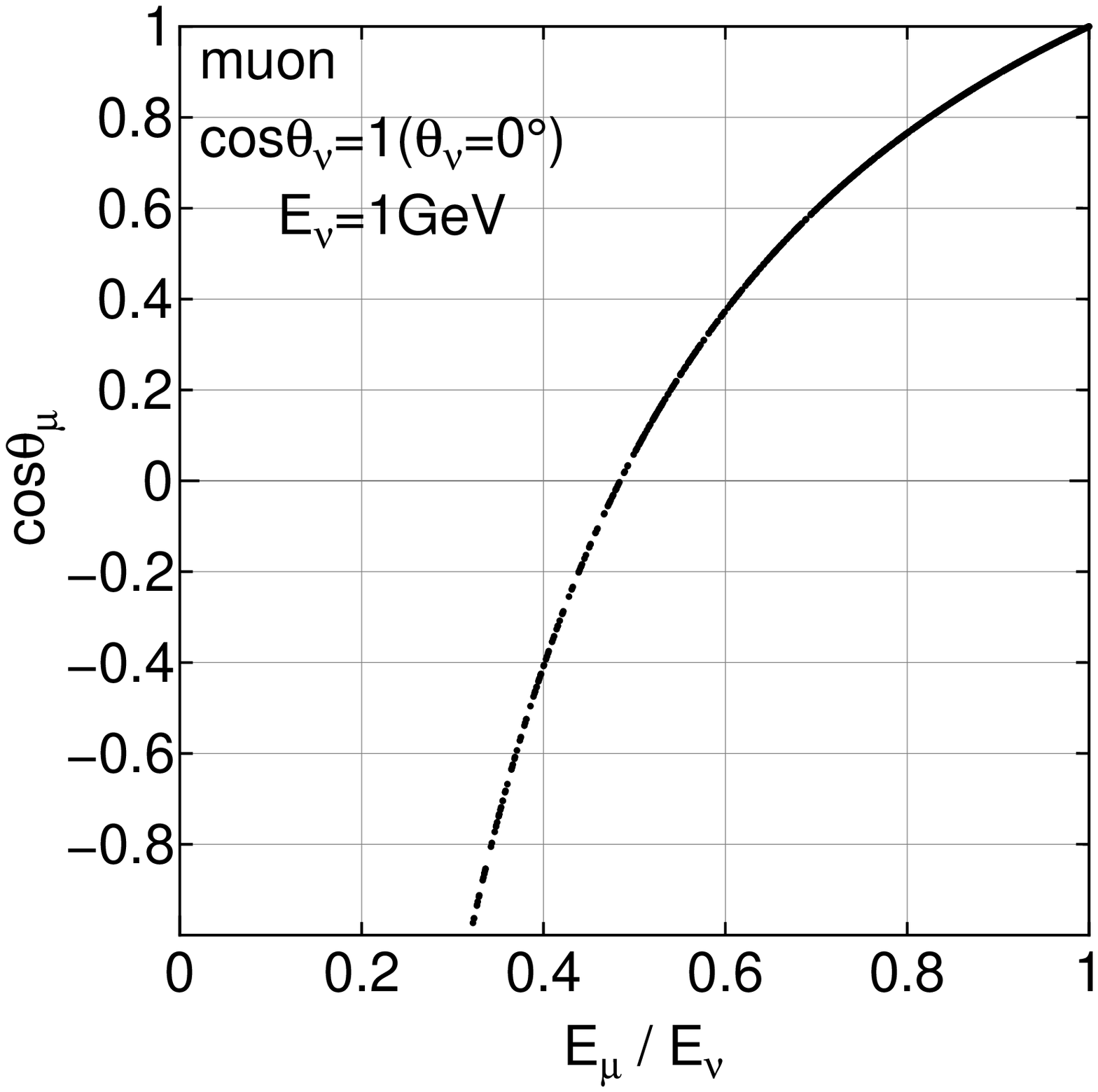}\hspace{1cm}
  \includegraphics{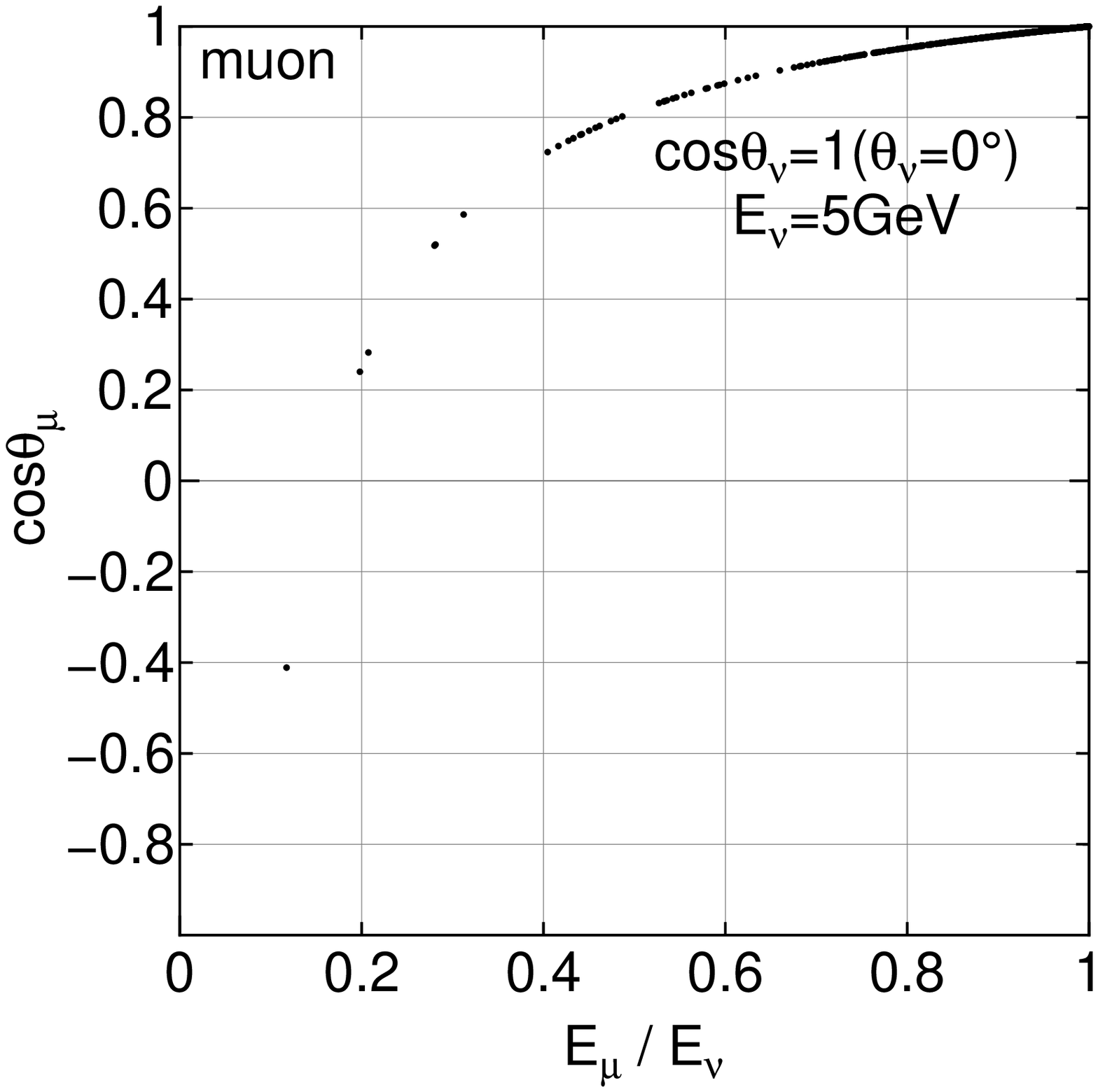}
}
\caption{
\label{figH004}
The scatter plots between the fractional energies of the produced muons 
and their zenith angles 
for vertically incident muon neutrinos with 0.5~GeV, 1~GeV and 5~GeV, 
respectively.
 The sampling number is 1000 for each case.
}
\end{center}
\vspace{0.5cm}
\hspace{2.5cm}(a)
\hspace{5.5cm}(b)
\hspace{5.5cm}(c)
\vspace{-0.3cm}
\begin{center}
\resizebox{\textwidth}{!}{%
  \includegraphics{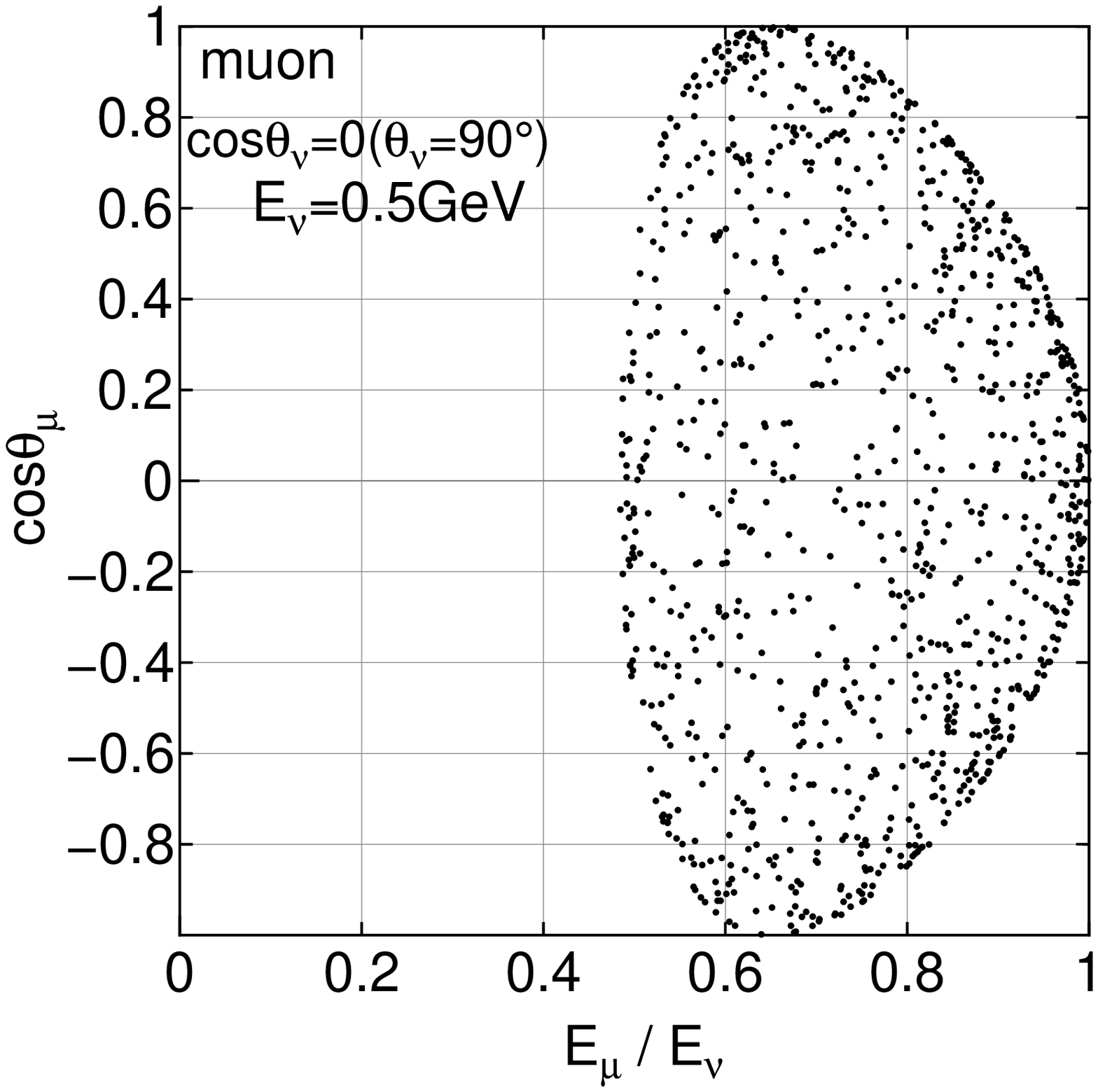}\hspace{1cm}
  \includegraphics{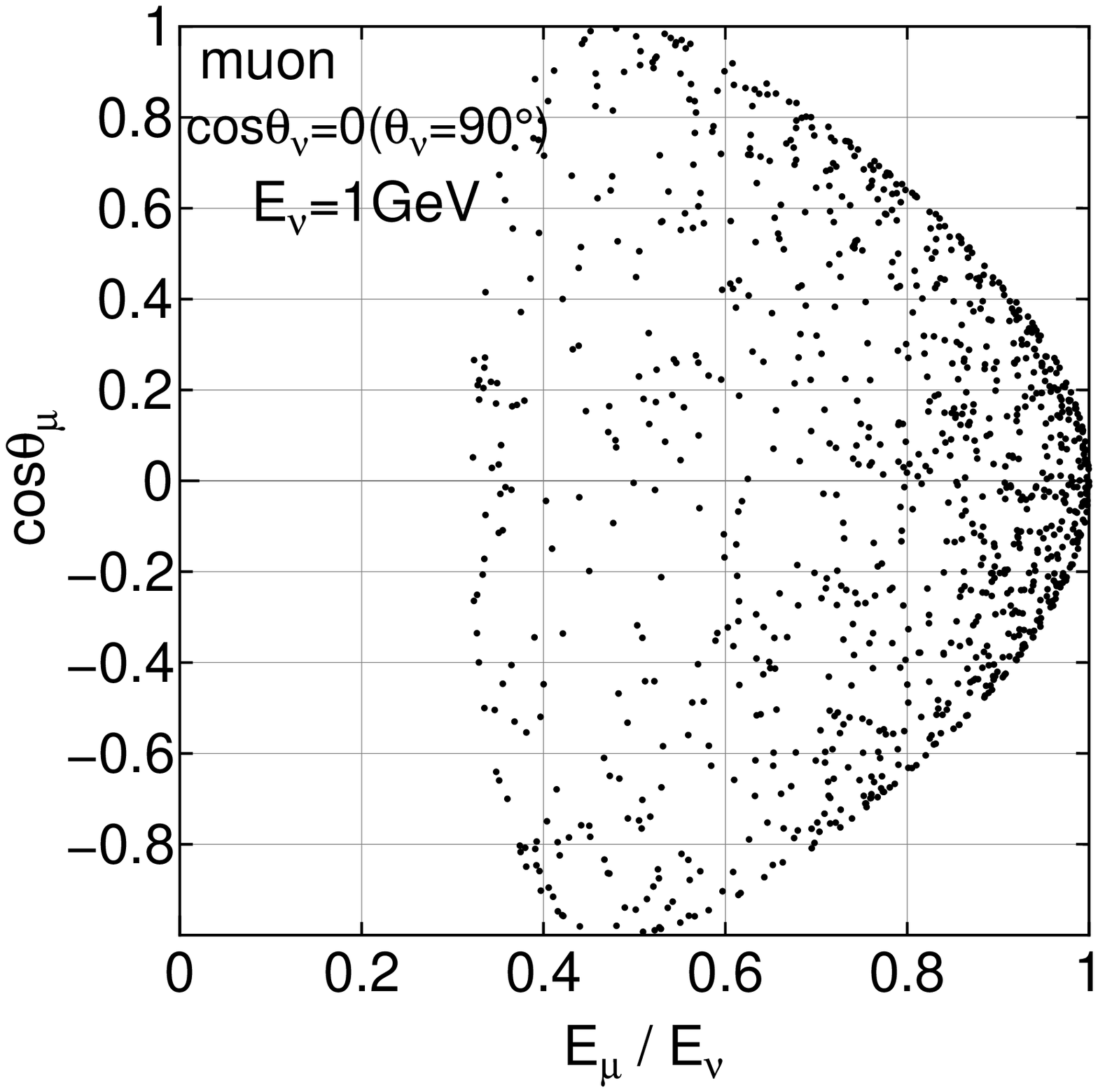}\hspace{1cm}
  \includegraphics{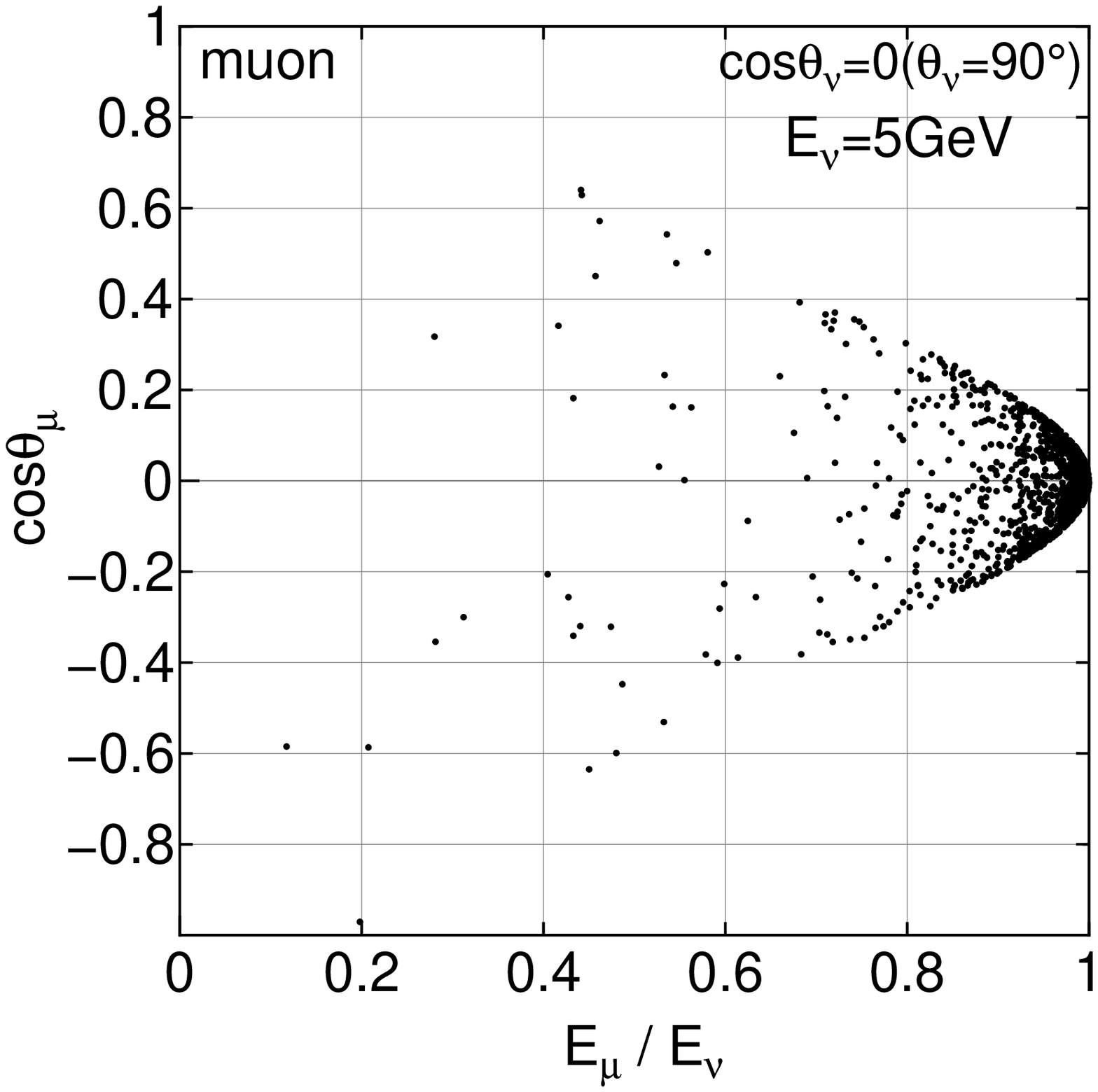}
}
\caption{
\label{figH005} 
The scatter plots between the fractional energies of the produced muons 
and their zenith angles 
for horizontally incident muon neutrinos with 0.5~GeV, 1~GeV and 5~GeV, 
respectively.
 The sampling number is 1000 for each case.
}
\end{center}
\vspace{0.5cm}
\hspace{2.5cm}(a)
\hspace{5.5cm}(b)
\hspace{5.5cm}(c)
\vspace{-0.3cm}
\begin{center}
\resizebox{\textwidth}{!}{%
  \includegraphics{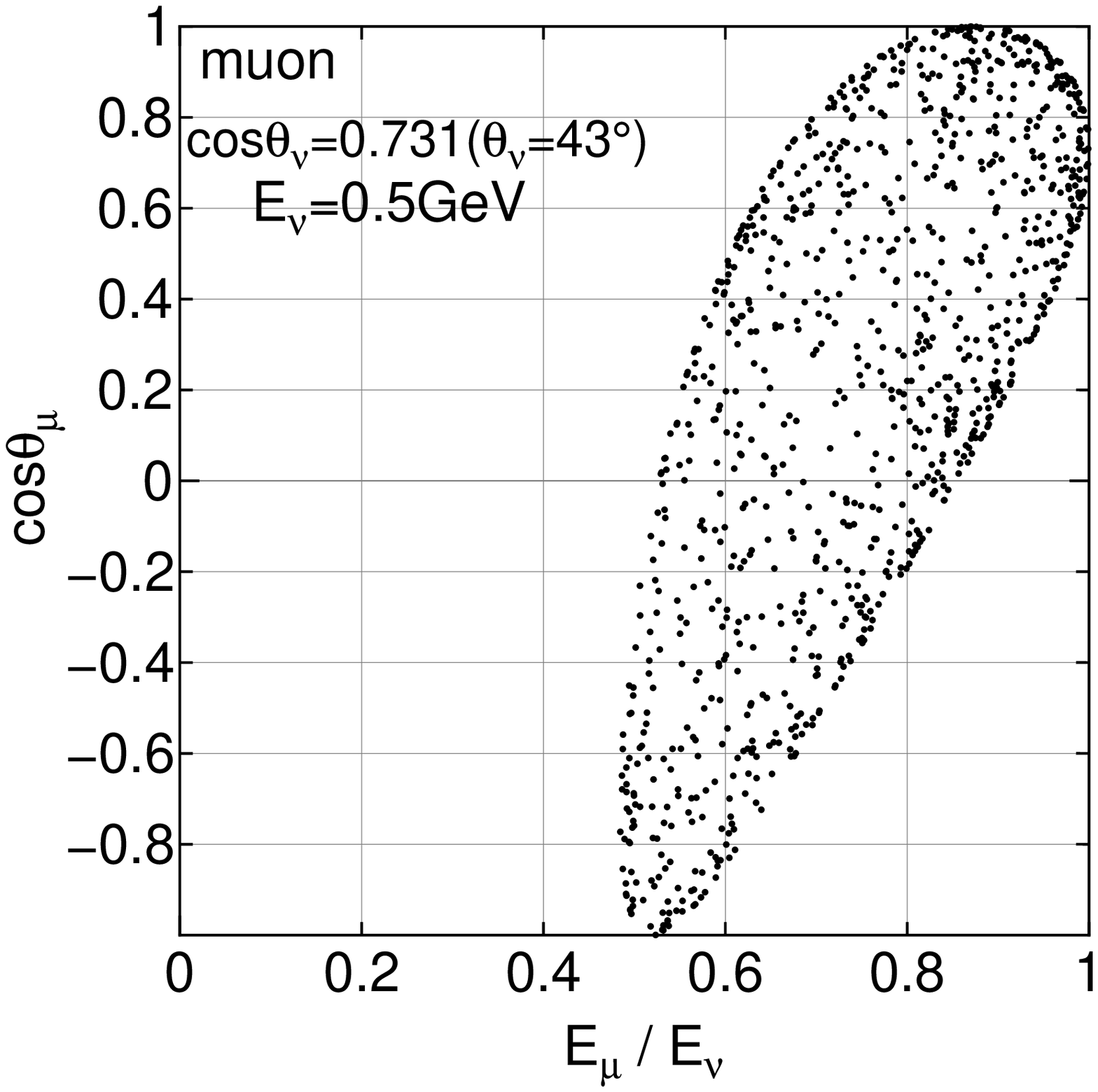}\hspace{1cm}
  \includegraphics{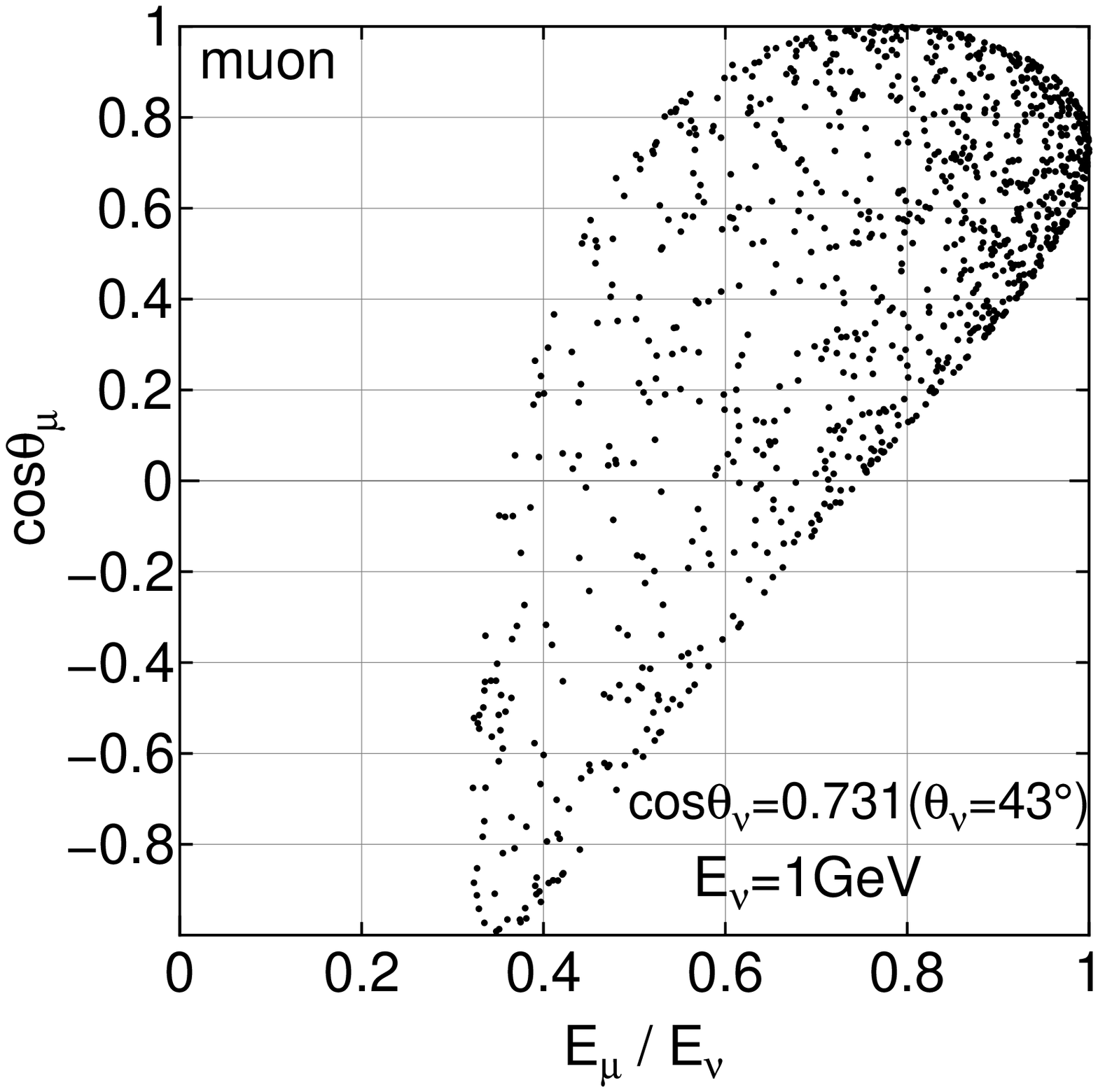}\hspace{1cm}
  \includegraphics{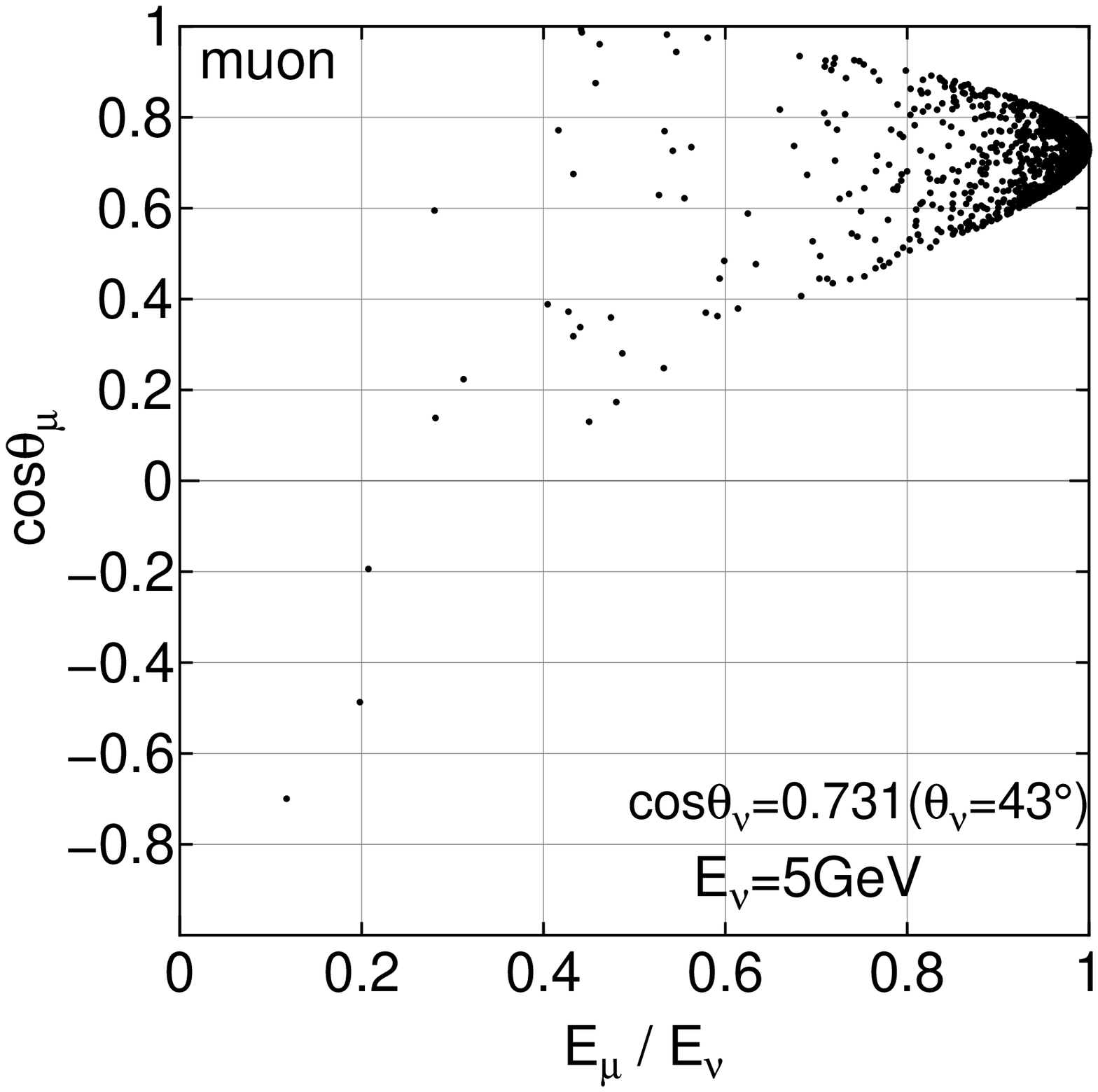}
  }
\caption{
\label{figH006} 
The scatter plots between the fractional energies of the produced muons 
and their zenith angles 
for diagonally incident muon neutrinos with 0.5~GeV, 1~GeV and 5~GeV, 
respectively.
 The sampling number is 1000 for each case.
}
\end{center}
\end{figure*} 

\begin{figure*}
\vspace{-1.0cm}
\hspace{2.5cm}(a)
\hspace{5.5cm}(b)
\hspace{5.5cm}(c)
\vspace{-0.3cm}
\begin{center}
\resizebox{\textwidth}{!}{%
  \includegraphics{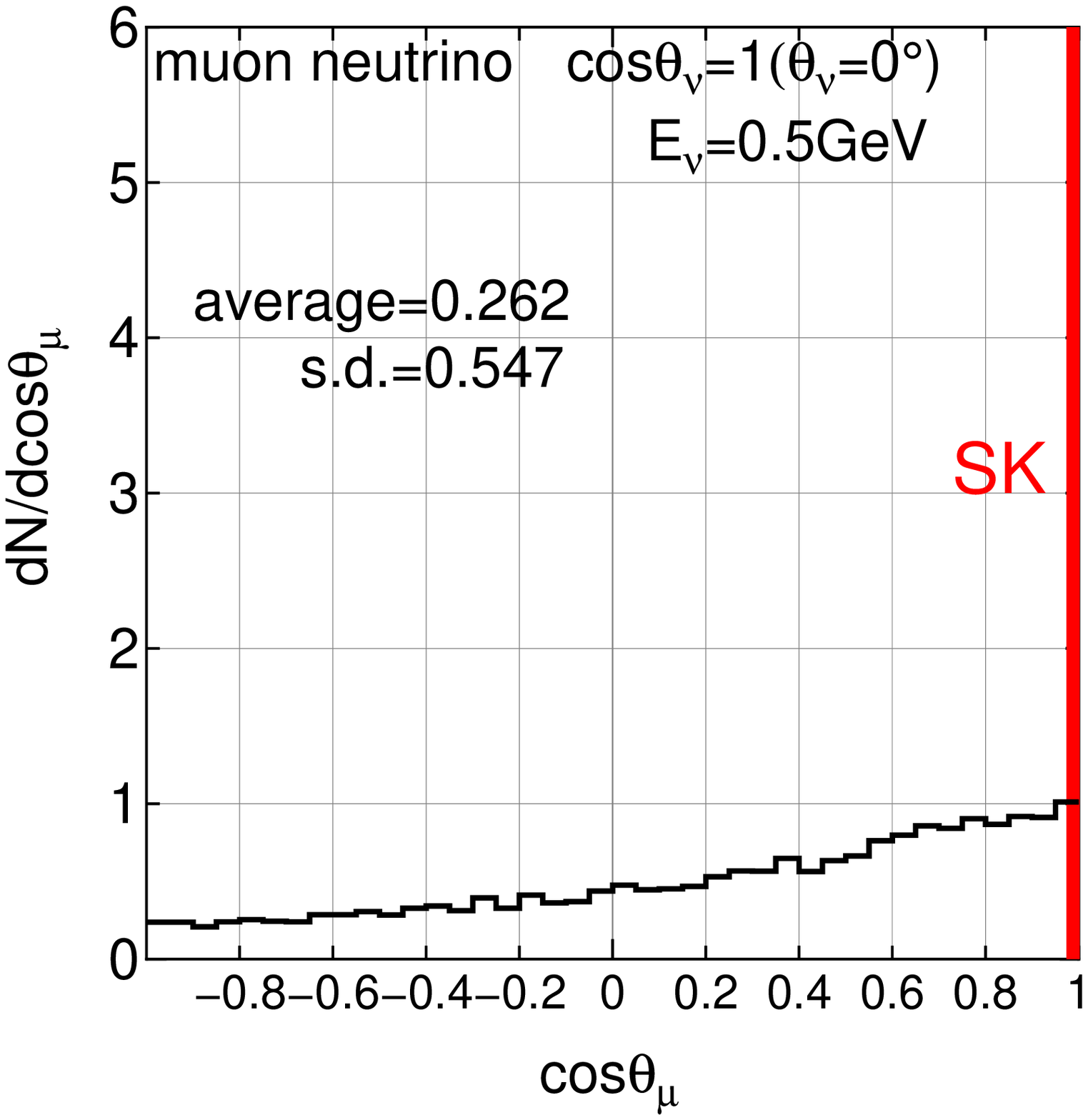}\hspace{1cm}
  \includegraphics{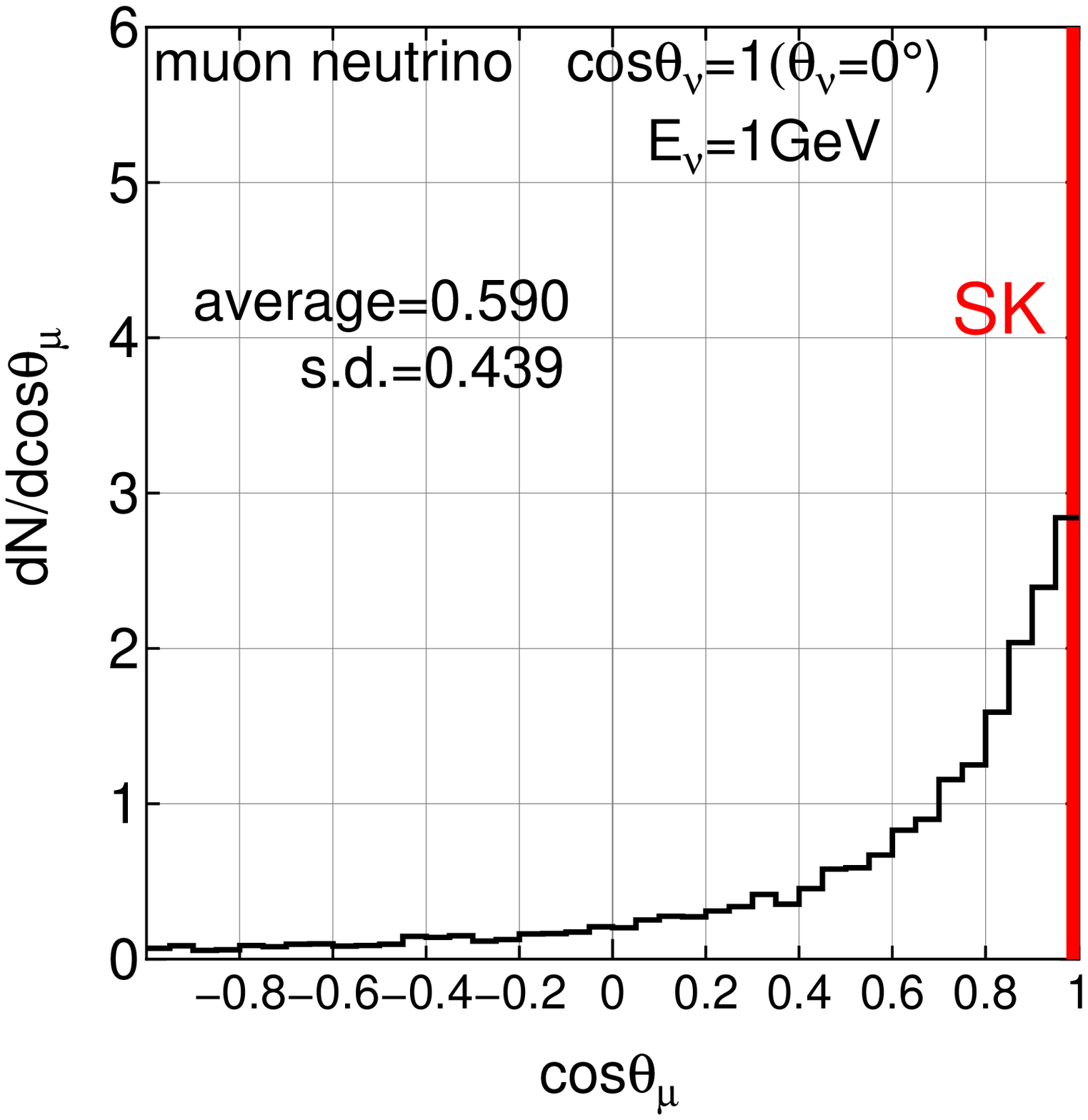}\hspace{1cm}
  \includegraphics{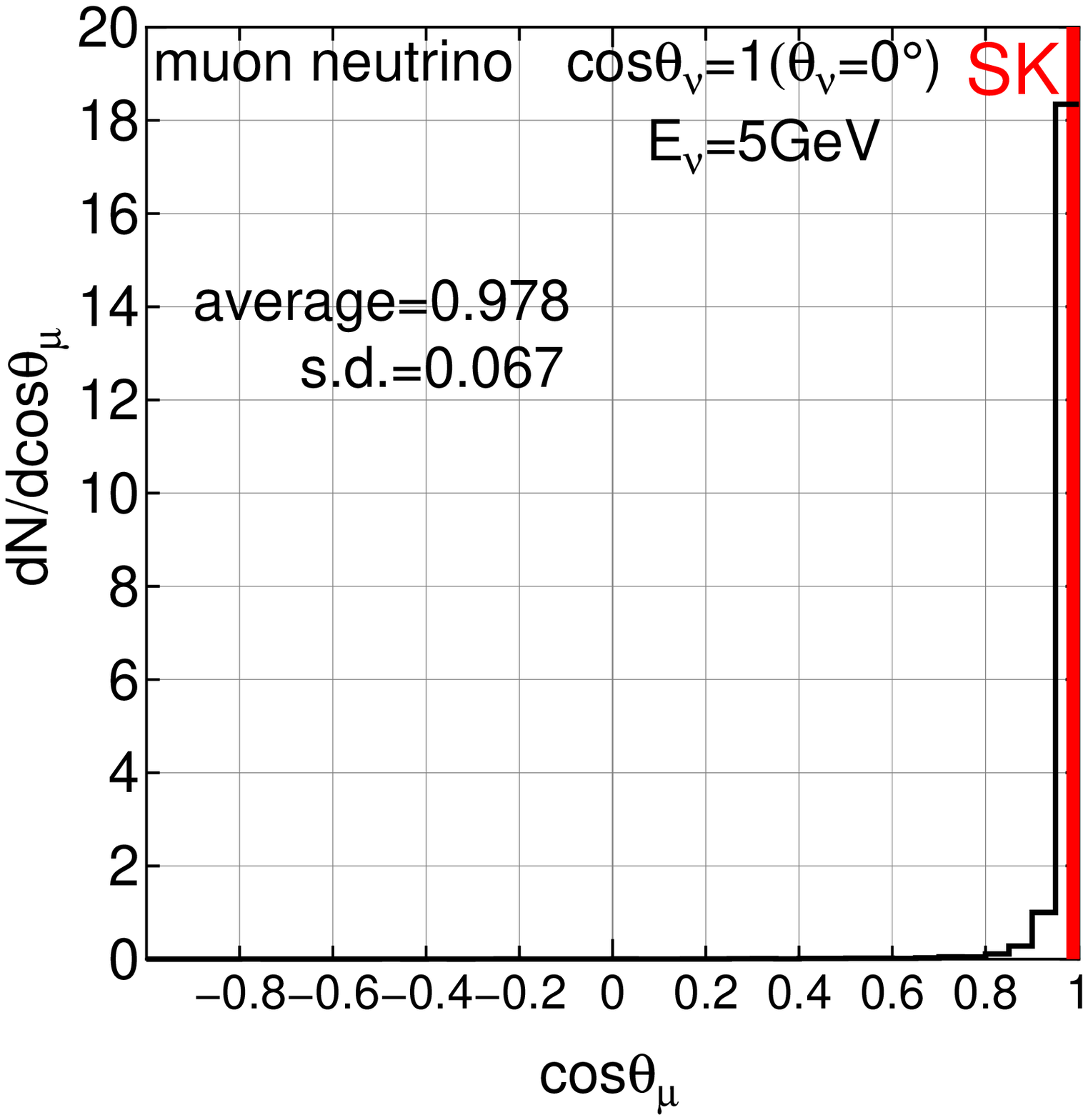}
  }
\caption{
\label{figH007} 
Zenith angle distribution of the muon for the vertically incident muon 
neutrino with 0.5~GeV, 1~GeV and 5~GeV, respectively. The sampling 
number is 10000 for each case.
SK stands for the corresponding ones under 
{\it the SK assumption on the direction}.
}
\end{center}
\vspace{0.5cm}

\hspace{2.5cm}(a)
\hspace{5.5cm}(b)
\hspace{5.5cm}(c)
\vspace{-0.3cm}
\begin{center}
\resizebox{\textwidth}{!}{%
 \includegraphics{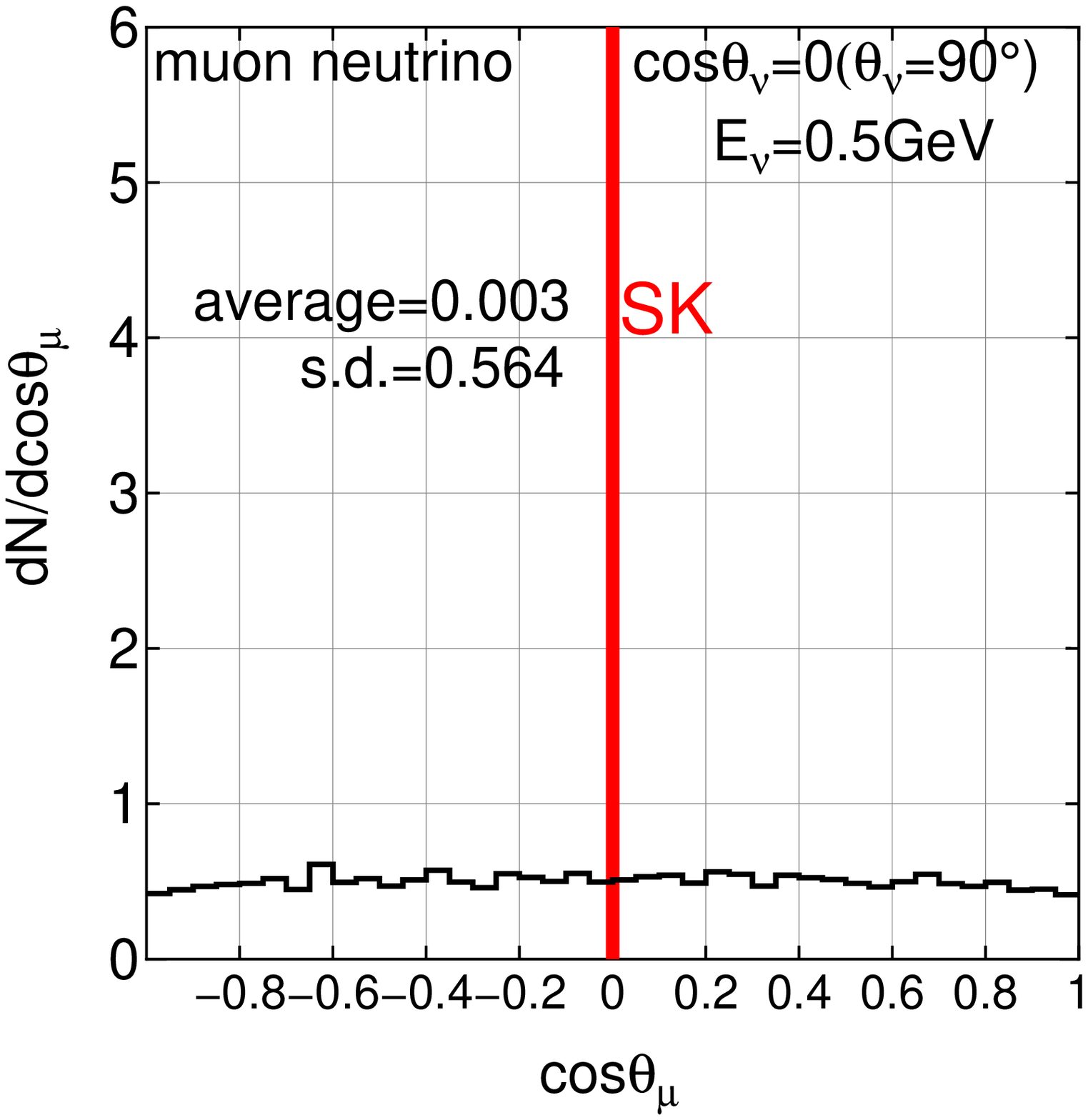}\hspace{1cm}
  \includegraphics{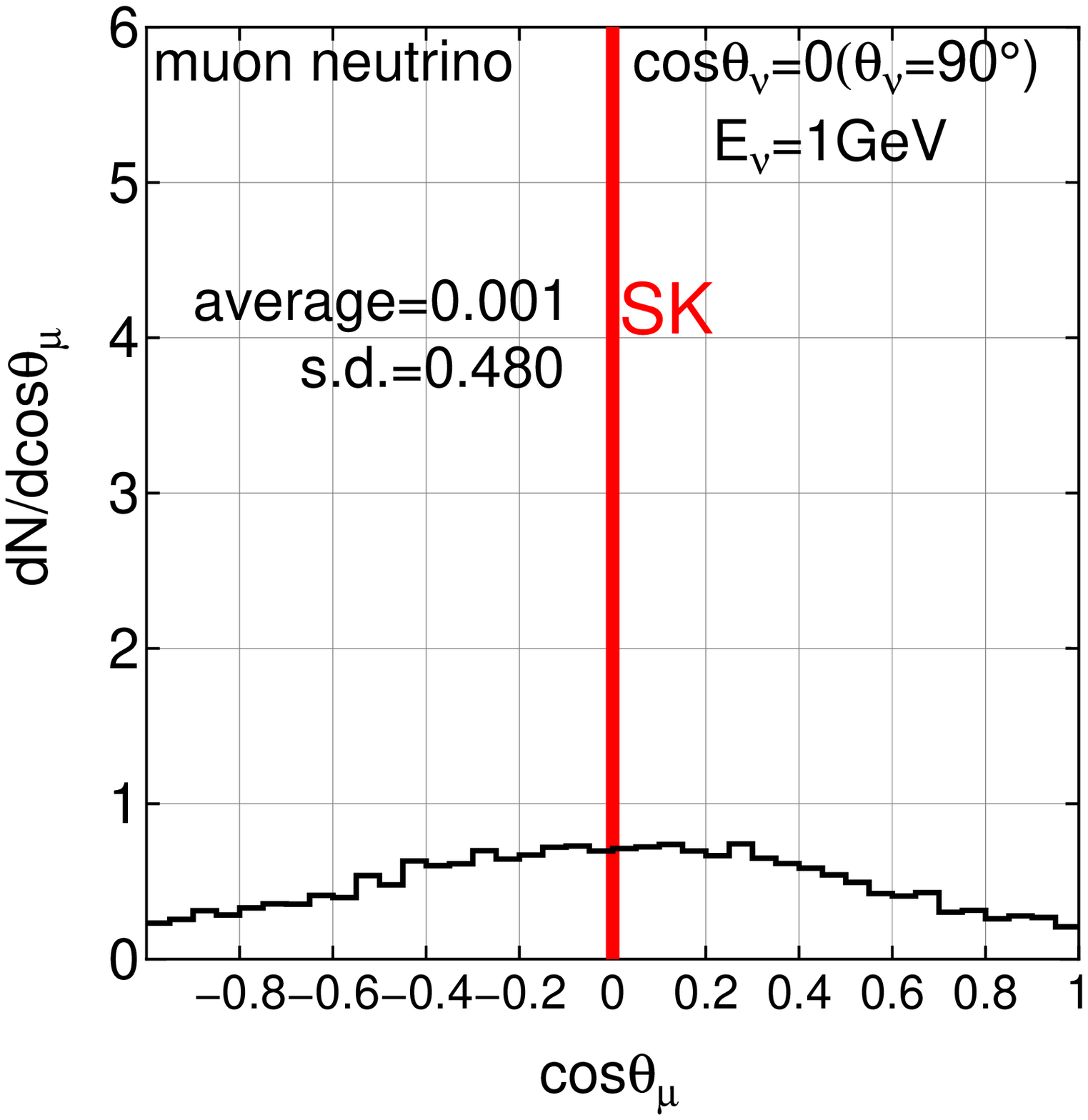}\hspace{1cm}
  \includegraphics{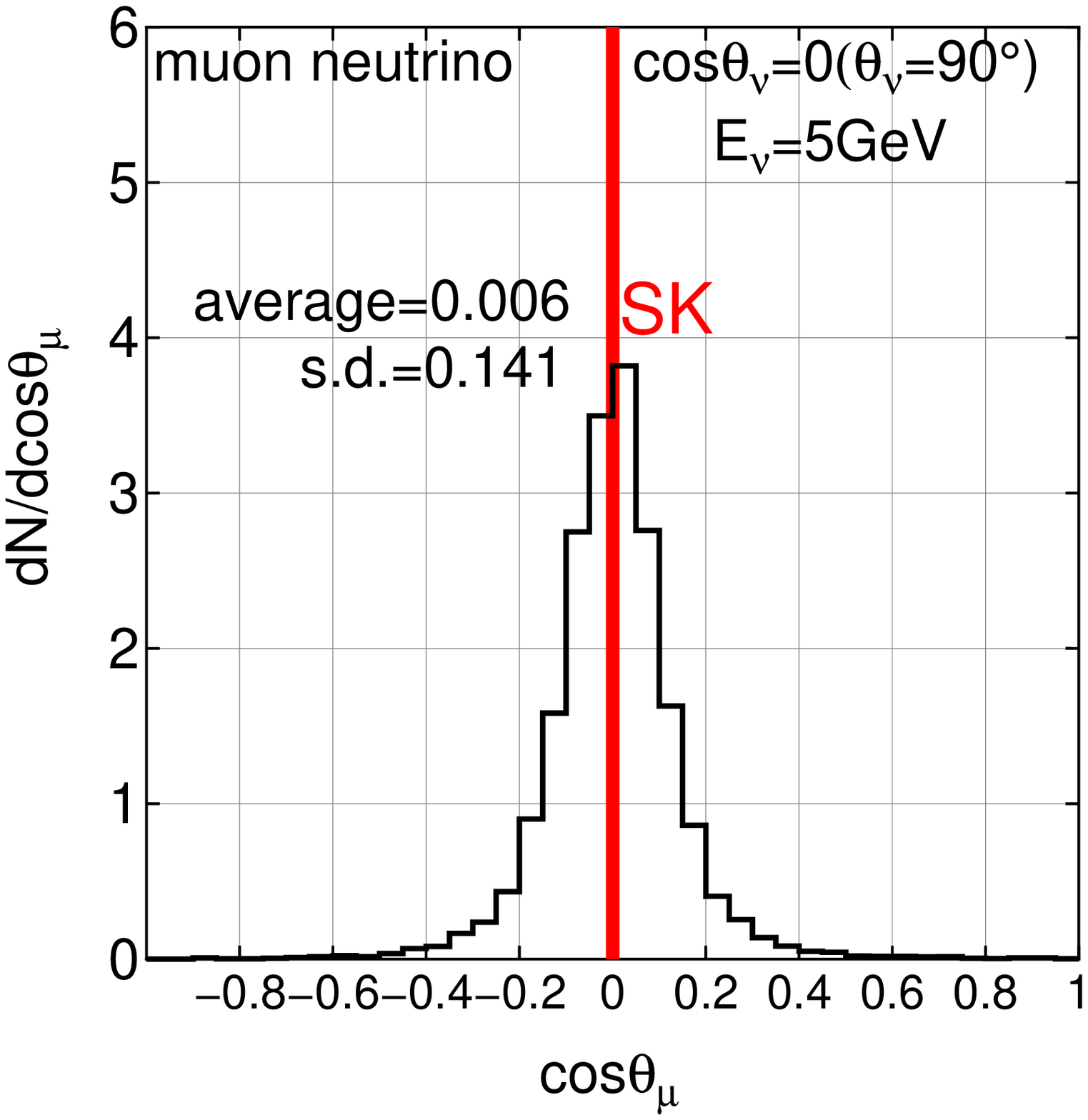}
  }
\caption{
\label{figH008} 
Zenith angle distribution of the muon for the horizontally incident muon 
neutrino with 0.5~GeV, 1~GeV and 5~GeV, respectively. The sampling number 
is 10000 for each case.
SK stands for the corresponding ones under 
{\it the SK assumption on the direction}.
}
\end{center}
\vspace{0.5cm}
\hspace{2.5cm}(a)
\hspace{5.5cm}(b)
\hspace{5.5cm}(c)
\vspace{-0.3cm}
\begin{center}
\resizebox{\textwidth}{!}{%
  \includegraphics{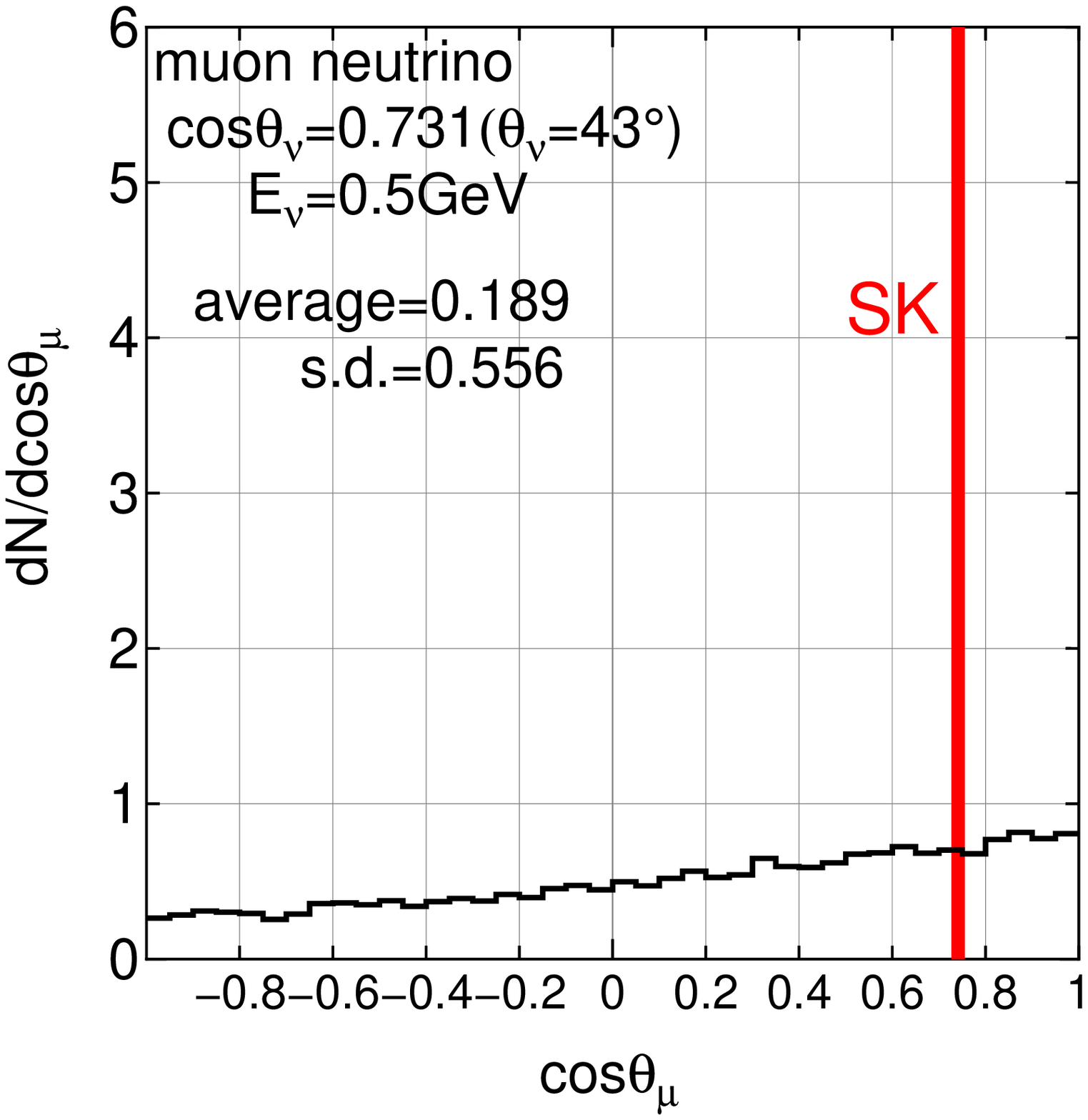}\hspace{1cm}
  \includegraphics{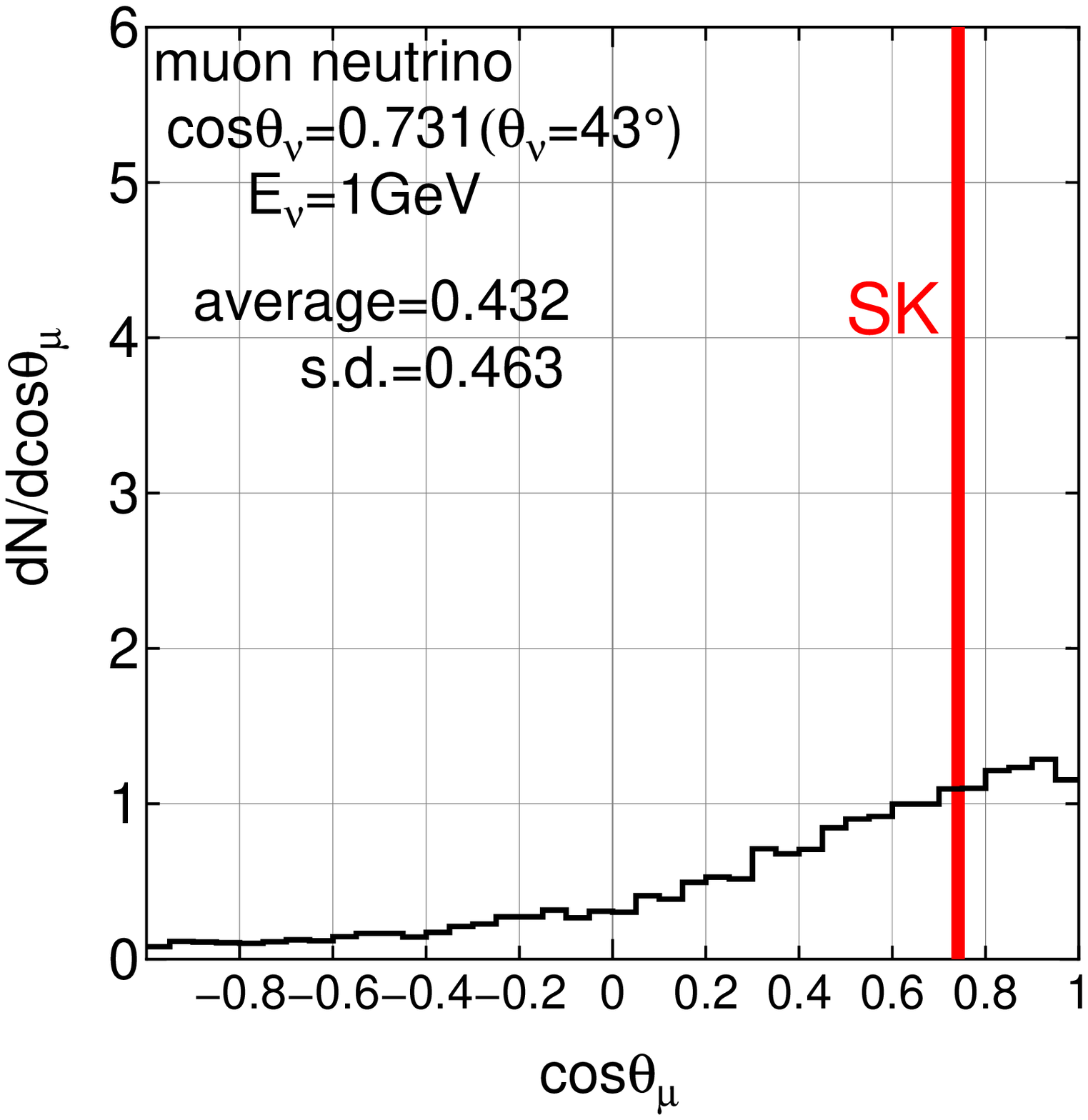}\hspace{1cm}
  \includegraphics{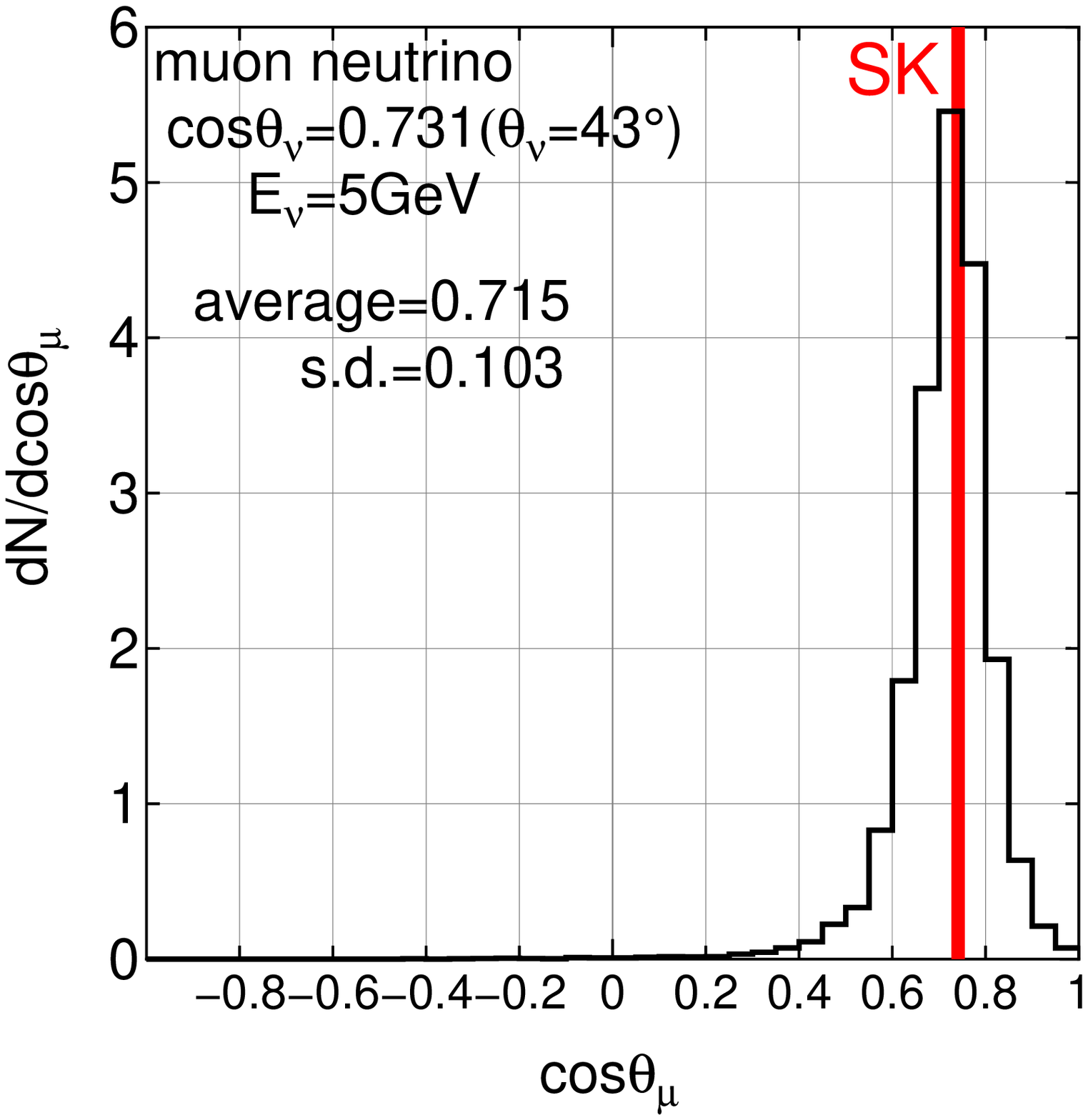}
  }
\caption{
\label{figH009} 
Zenith angle distribution of the muon for the diagonally incident muon 
neutrino with 0.5~GeV, 1~GeV and 5~GeV, respectively. The sampling 
number is 10000 for each case.
SK stands for the corresponding ones under 
{\it the SK assumption on the direction}.
}
\end{center}
\end{figure*} 
\subsection{Dependence of the spread of the zenith angle for emitted 
leptons on the energy 
of emitted leptons for different incident directions 
of the neutrinos with different 
energies}
  The detailed procedure for our Monte Carlo simulation is described in 
 Appendix A. 
 We give  the scatter plots between the fractional energy of the emitted 
muons and their zenith angle
 for a definite zenith angle of the incident neutrino with different 
energies in Figures~\ref{figH004} to \ref{figH006}.
 In Figure~\ref{figH004}, we give 
the scatter plots for vertically incident neutrinos with different 
energies 
0.5, 1 and 5~GeV. In this case, the relations between the emitted 
energies of the muons and their zenith angles are unique, which comes 
from the definition of the zenith angle of the emitted lepton. However, 
the densities (frequencies of the event number) along each curves are 
different 
in position to position and depend on the energies of the incident 
neutrinos. Generally speaking, densities along the curves become greater 
toward  $\cos\theta_{\mu(\bar{\mu})}= 1$. In this case, 
$\cos\theta_{\mu(\bar{\mu})}$ is never influenced by the azimuthal angle 
in the scattering by the 
definition\footnote{
The zenith angles of the particles concerned are measured from 
the vertical upward direction.}.
 
On the contrast, it is shown in Figure~\ref{figH005} 
that the horizontally incident 
neutrinos give the widest zenith angle distribution for the 
emitted muons with the same fractional energies
 due to the effect of the azimuthal angles.
 The lower the energies of the incident neutrinos are, the  
wider the spreads of the scattering angles of emitted muons
$\theta_{\mu}$ become,
which leads to wider zenith angle distributions for the emitted muons.
  As easily understood from Figure~\ref{figH006}, 
the diagonally incident neutrinos give the intermediate zenith angle 
distributions for the emitted  muons between those for vertically 
incident neutrinos and those for horizontally incident neutrinos.  

It should be noticed from the figures that the influence of the 
azimuthal angle in QEL over the cosines of the zenith angle for the
 incident neutrino becomes effective in the more inclined neutrino,
 even if the scattering angle due to QEL is not so large.
  The effect of the azimutal angle in QEL is not taken into account 
in the Monte Carlo simulation by Super-Kamiokande Collaboration,
 because their Monte Carlo simulation is a detector simulation and 
this effect cannot be taken into account by their nature.

\subsection{Zenith angle distribution 
of the emitted leptons for different incident directions 
of the neutrinos with different 
energies}
%
In Figures~\ref{figH007} to \ref{figH009}, 
we express Figures~\ref{figH004} to \ref{figH006} in a different way. 
We sum up muon events with different emitted energies for given 
zenith angles. As the result of it, we obtain frequency 
distribution of the neutrino events as  a function of 
$cos\theta_{\mu}$ for 
different incident directions and different incident energies of 
neutrinos.

In Figures~\ref{figH007}(a) to \ref{figH007}(c), we give the zenith angle 
distributions of the emitted muons for the case of vertically incident 
neutrinos with different energies, say,
 $E_{\nu}=$ 0.5, 1 and 5~GeV.

Comparing the case for 0.5 GeV with that for 5 GeV, we understand the big 
contrast between them as for the zenith angle distribution. The scattering 
angle of the emitted muon for 5 GeV neutrino is relatively small (See, 
Table 1), so that the emitted muons keep roughly the same direction as 
their 
original neutrinos. In this case, the effect of their azimuthal angle on 
the zenith angle is also smaller. However, in the case for 0.5 GeV 
which is the dominant energy for single ring muon events
in the Super-Kamiokande, there is even 
a possibility for the emitted muon to be emitted in the backward direction 
due to the larger angle scattering, the effect of which is enhanced by 
their azimuthal angle.

The most frequent occurrence in the backward direction of the emitted 
muon appears in the horizontally incident neutrino as shown in Figs. 8(a) 
to 8(c). In this case, the zenith angle distribution of the emitted muon 
should be symmetrical with regard
to the horizontal direction. Comparing the case for 
5 GeV with those both for $\sim$0.5 GeV and for $\sim$1 GeV, even 1 GeV 
incident neutrinos lose almost the original sense of the incidence if we 
measure it by the zenith angle of the emitted muon. 
Figures~\ref{figH009}(a) to \ref{figH009}(c) for 
the diagonally incident neutrinos tell us that the situation for diagonal 
case lies between the case for the vertically incident neutrinos and that 
for horizontally incident ones.
  SK in the figures denotes {\it the SK assumption on the direction} of 
incident neutrinos. 
From the Figures~\ref{figH007}(a) to \ref{figH009}(c), it is clear that the scattering 
angles of emitted muons influence their zenith angles, which is enhanced 
by their azimuthal angles, particularly for
more inclined directions of the incident neutrinos. 

\begin{figure*}
\vspace{-0.5cm}
\hspace{2cm}(a)
\hspace{5cm}(b)
\hspace{5cm}(c)
\vspace{-0.2cm}
\begin{center}
\resizebox{\textwidth}{!}{%
  \includegraphics{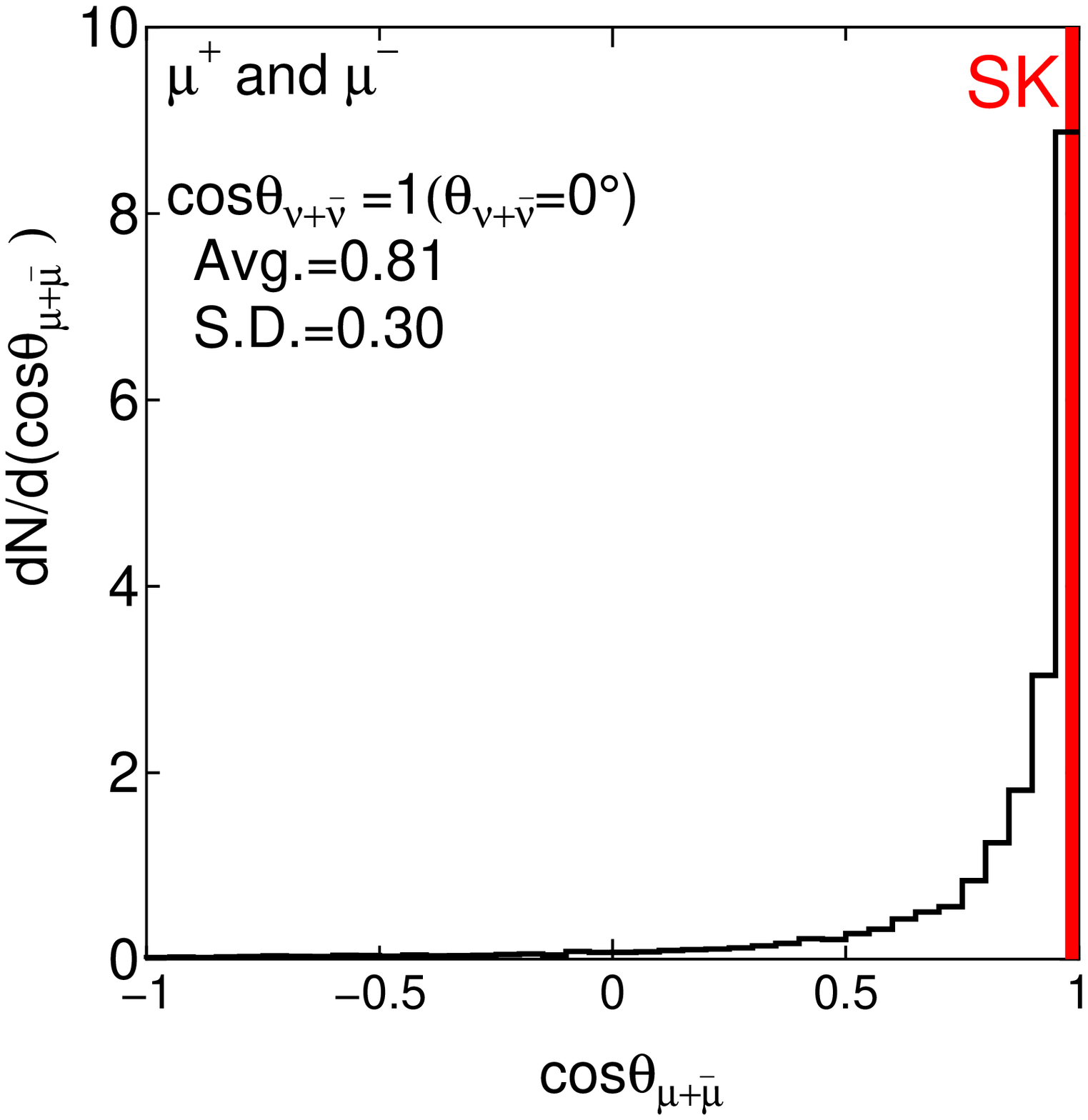}\hspace{1cm}
  \includegraphics{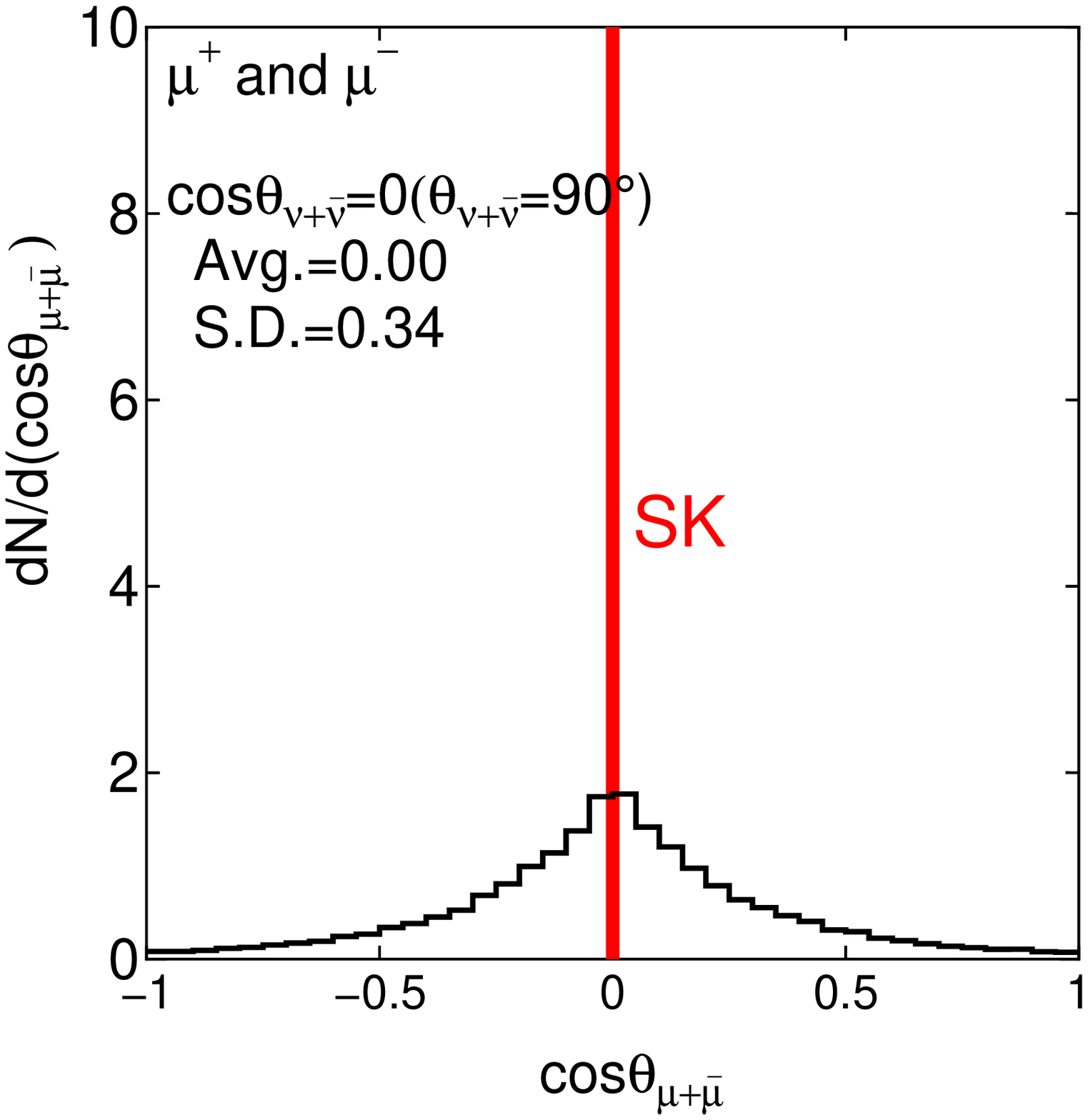}\hspace{1cm}
  \includegraphics{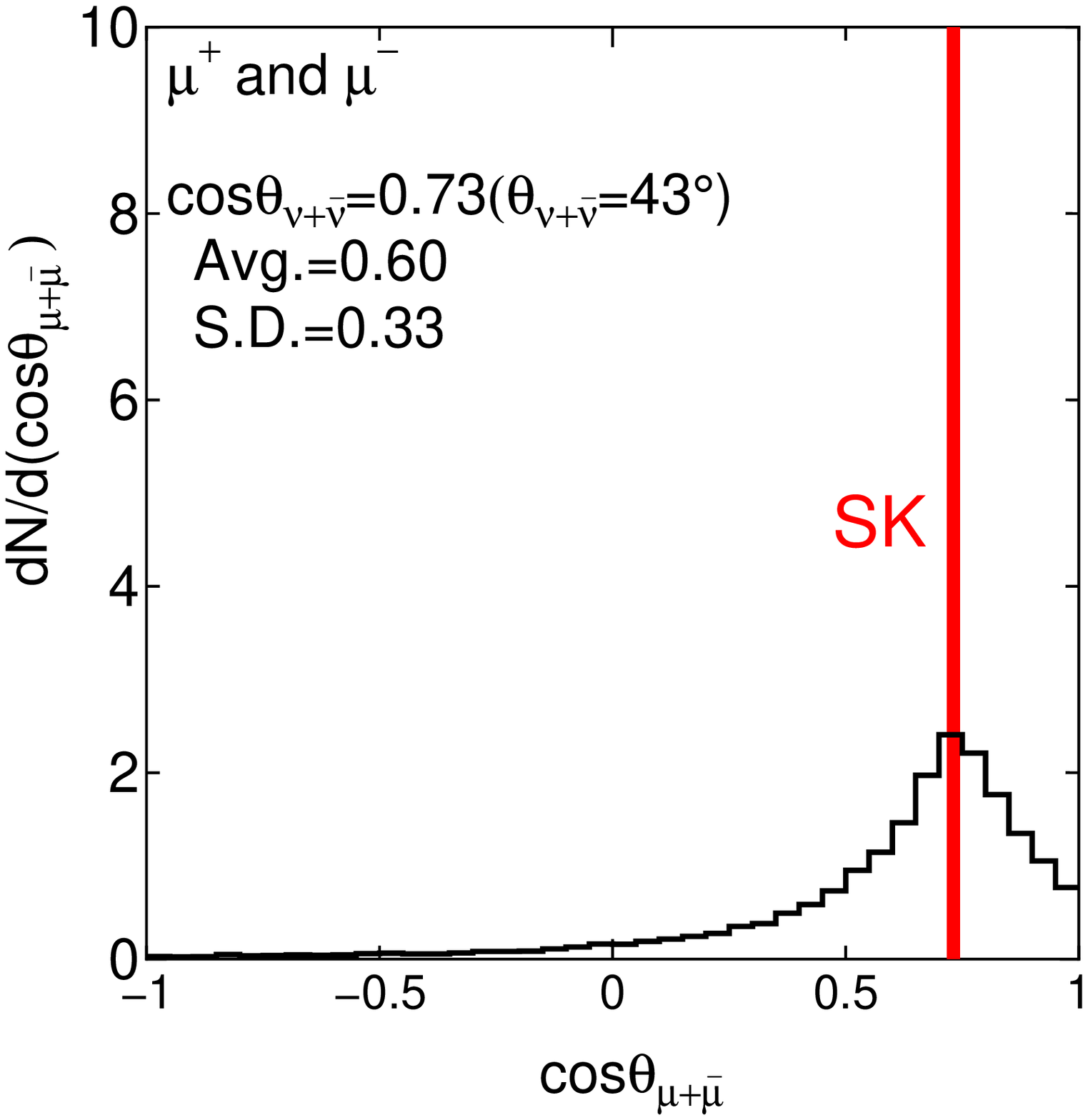}\hspace{1cm}
  }\\
 \end{center}
\caption{\label{fig:10} 
Zenith angle distribution of $\mu^-$ and $\mu^+$ for  $\nu$  and   
$\bar{\nu}$ for the incident neutrinos with the vertical, horizontal and 
diagonal directions, respectively (see Figure~3). 
The overall neutrino spectra at Kamioka site are taken into account.
 The sampling number is 10000 for each 
case. SK stand for the corresponding ones under 
{\it the SK assumption on the direction}.
}
\end{figure*} 
%
\subsection{Zenith Angle Distribution of Fully Contained Events
for a Given Zenith Angle of the Incident Neutrino, Taking Their Energy 
Spectrum into Account}

In the previous sections, we discuss the relation between the zenith angle 
distribution of the incident neutrino with a definite energy and that of 
the emited muons produced by the neutrino for the different incident 
direction. 
In order to apply our inspection around the uncertainty of 
\textit{the SK assumption on the direction} for \textit{Fully Contained 
Events} 
to the real situation,
 we must consider the 
effect of the energy spectrum of the incident neutrino. The Monte Carlo 
simulation procedures for this purpose are given in Appendix B.

In Fig. \ref{fig:10}, we give the zenith angle distributions of the sum of 
$\mu^+(\bar{\mu})$ and $\mu^-$ for a given zenith angle of 
$\bar{\nu}_{\bar{\mu}}$ and $\nu_{\mu}$, taking into account
different primary neutrino energy spectra for different directions
at Kamioka site.
SK in the figures denotes {\it the SK assumption on the direction}. 
From the figures, it is clear that 
{\it the SK assumption on the direction} does not hold.
 Namely, we can 
conclude that the scattering angle of the emitted muons 
acompanied by 
their azimuthal angles influence their zenith angle distribution for 
 all directions. 

%
\section{
Super-Kamiokande Assumption on the Direction and the Real Relation 
between
$cos\theta_{\nu(\bar{\nu})}$ and $cos\theta_{\mu(\bar{\mu})}$ 
and the corresponding relation between 
 $L_{\nu(\bar{\nu})}$ and $L_{\mu(\bar{\mu})}$ 
} 


\subsection{Correlation between 
${\cos\theta_{\nu}}$ and ${\cos{\theta}_{\mu}}$
}

Now, we extend the results for the definite zenith angle obtained in the 
previous sections to the case in which we consider the zenith angle 
distribution of the incident neutrinos totally.

Here, we examine the real correlation between ${\cos\theta_{\nu}}$ and 
${\cos{\theta}_{\mu}}$, by performing the exact Monte Carlo simulation. 
 

The detail for the simulation procedure is given in Appendix C.

In order to obtain the zenith angle distribution of the emitted 
leptons for that of the incident neutrinos, we divide 
the range of cosine 
of the zenith angle of the incident neutrino into twenty regular 
intervals from $\cos{\theta_{\nu}}=0$ to $\cos{\theta_{\nu}}=1$.
For the given interval of $\cos{\theta_{\nu}}$, we carry out the 
exact Monte Carlo simulation, and obtain the cosine of the zenith 
angle of the emitted leptons.
 
Thus, for each interval of $\cos{\theta_{\nu}}$, we obtain the
corresponding zenith angle distribution of the emitted leptons. 
Then, we sum up these corresponding ones over all zenith angles of 
the incident neutrinos and we finally obtain the relation between 
the zenith angle distribution for the incident neutrinos and that 
for the emitted leptons. 

In a similar manner, we could obtain between $\cos{\theta_{\bar{\nu}}}$
and $\cos{\theta_{\bar{\mu}}}$ for anti-neutrinos. The situation 
for anti-neutrinos is essentially the same as that for neutrinos.

\begin{figure*}
\begin{center}
\vspace{-1.5cm}
\rotatebox{90}{%
\resizebox{0.7\textwidth}{!}{%
  \includegraphics{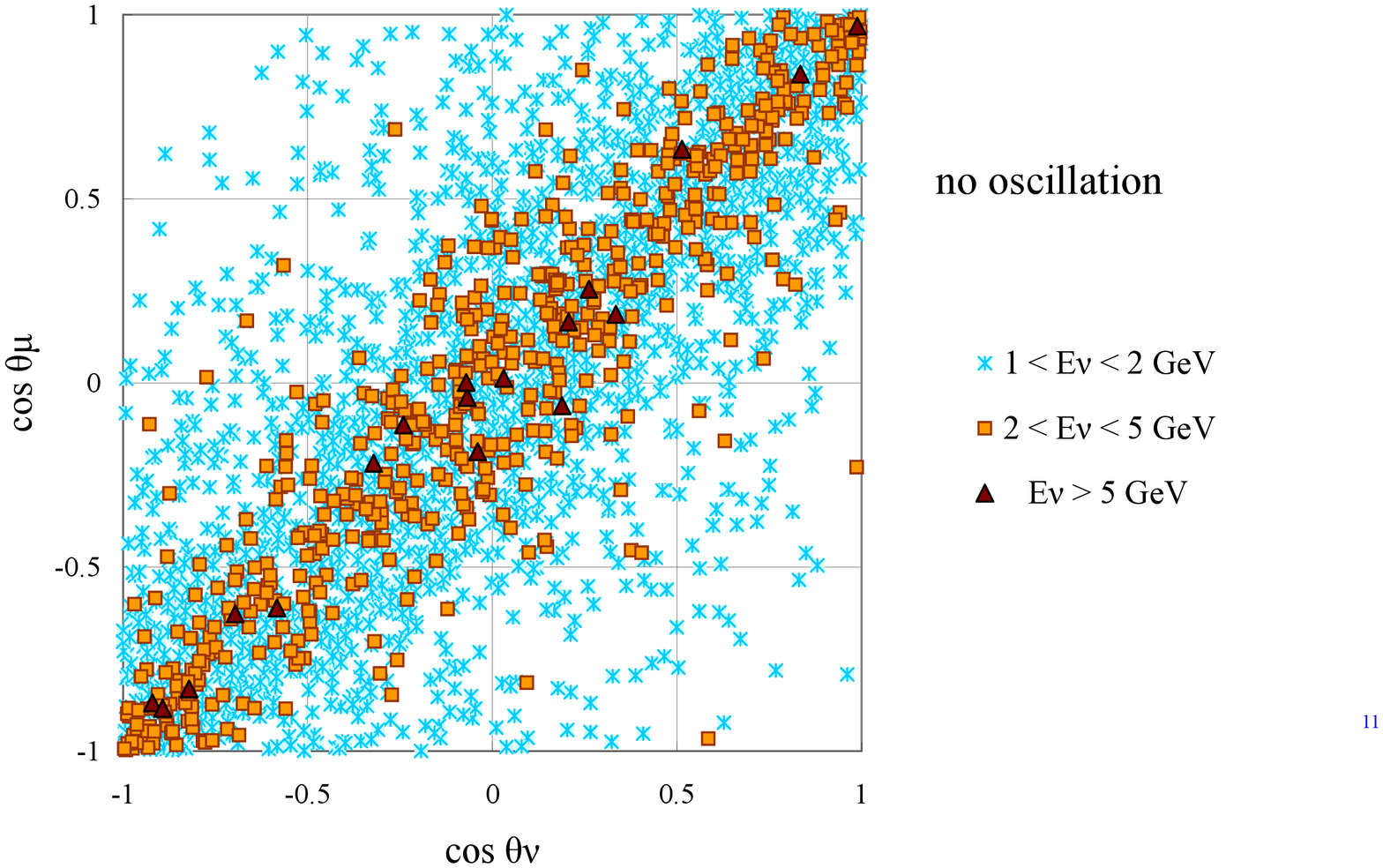}
  }}
\vspace{-1.5cm}
\caption{ Correlation diagram between $\cos{\theta}_{\nu}$ and $\cos{\theta}_{\mu}$ for null oscillation
for different neutrino energy regions.}
\label{figR011} 
\vspace{-0.2cm}
\rotatebox{90}{%
\resizebox{0.7\textwidth}{!}{%
  \includegraphics{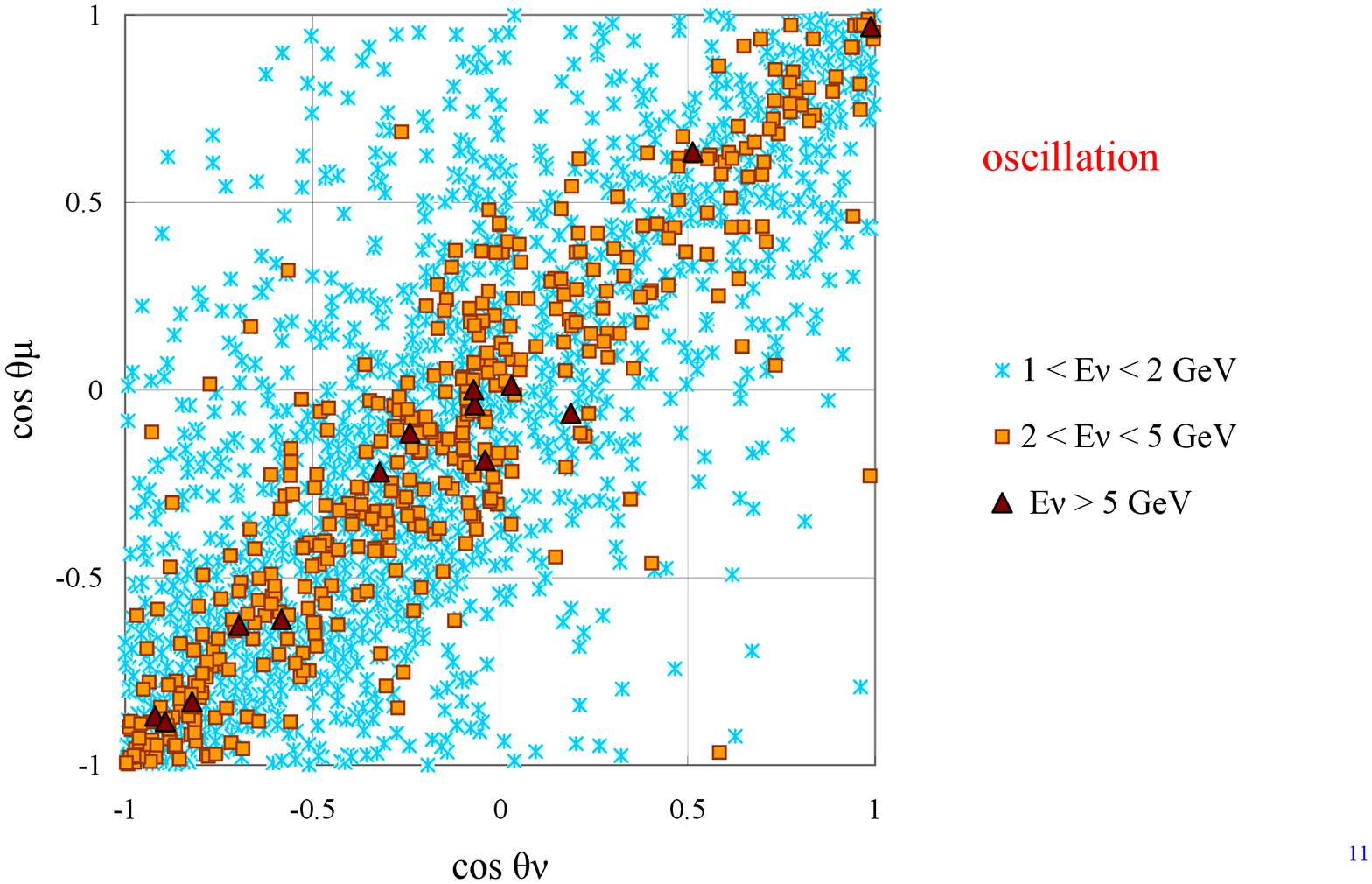}
  }}
\vspace{-1.5cm}
\caption{ Correlation diagram between $\cos{\theta}_{\nu}$ and $\cos{\theta}_{\mu}$ for oscillation
for different neutrino energy regions.}
\label{figR012}
\end{center}
\end{figure*} 
\begin{figure*}
\begin{center}
\vspace{-1.5cm}
\rotatebox{90}{%
\resizebox{0.7\textwidth}{!}{%
  \includegraphics{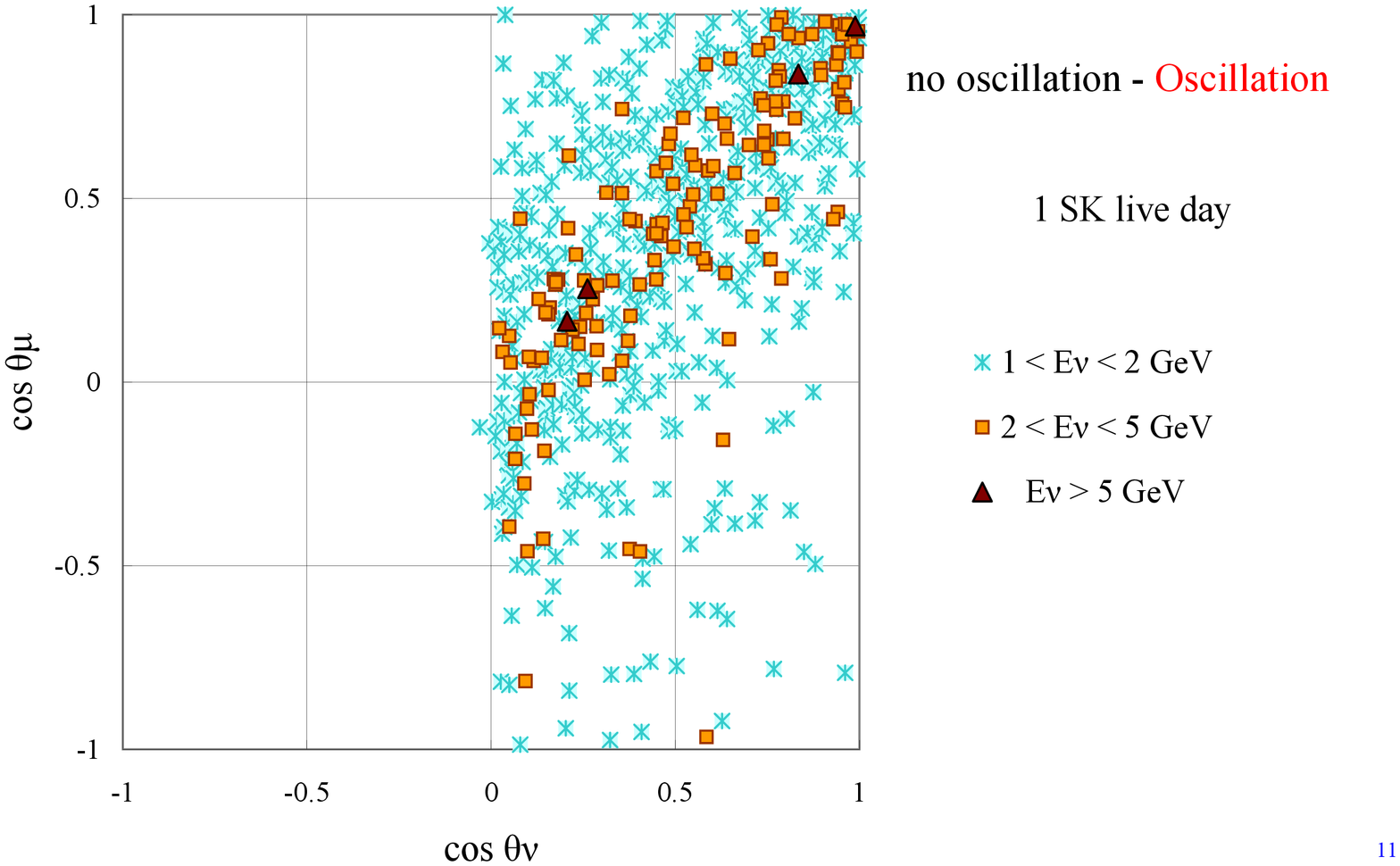}
  }}
\vspace{-1.5cm}
\caption{ Correlation diagram between $\cos{\theta}_{\nu}$ and $\cos{\theta}_{\mu}$ for the events which exist in null oscillation
but disappear due to oscillation for different neutrino energy regions.}
\label{figR013} 

\end{center}
\end{figure*} 

 Super-Kamiokande Collaboration have made the comprehensive analysis 
of the all possible types of the neutrino events on the neutrino
 oscillations \cite{Ashie2} for 1489.2 live days. 
We restrict our analysis to the neutrino events due to QEL to make 
uncertainties around the interpretation small as possible.

In Figure~\ref{figR011}, we give the correlation diagram between 
${\cos\theta_{\nu}}$ and ${\cos\theta_{\mu}}$ for 1489.2 
live days without oscillation.
Here, we classify the correlations with regard to different 
incident neutrino energies.
 It is clear from Figure~\ref{figR011} that the 
neutrino events for $E_{\nu}>$ 5~GeV exist on the line 
${\cos\theta_{\nu}} = {\cos\theta_{\mu}}$ roughly, 
and the neutrino events for 2~$<E_{\nu}<$~5~GeV exist along
the line ${\cos\theta_{\nu}} = {\cos\theta_{\mu}}$
with explicit deviations. 
Namely, in higher neutrino energy regions, 
{\it the SK assumption on the direction} 
roughly holds.
However, if we pay our attention to lower energies,
for 1~$<E_{\nu}<$~2~GeV, we easily understand that
their distribution deviates largely from 
the line ${\cos\theta_{\nu}} = {\cos\theta_{\mu}}$, 
showing clearly that {\it the SK assumption on the direction}
does not hold. 

 Consequently, in sum, we can conclude that
{\it the SK assumption on the direction}
 does not hold as a whole, 
taking account of the fact that lower energy neutrino events
 occupy the majority in the detector due to the strong steepness 
of the incident neutrino energy spectra. 
In Figure~\ref{figR012}, we give the correlation diagram between 
${\cos\theta_{\nu}}$ and ${\cos\theta_{\mu}}$ for 1489.2 
live days with oscillation.
The Monte Carlo simulation with oscillation is carried out with
the use of the rejection method with regard to the survival 
probability for a given flavor, which is based on the results
obtained by the Monte Carlo simulation without oscillation
\footnote{
We can carry out the Monte Carlo simulation for the case with oscillation
quite independently from that without oscillation, of course. Howerver,
the adoption of the rejection method make clearer the effect of
the survival probability for the given flavor.
}.

The regions in the correlation diagrams 
in Figures~\ref{figR011} and \ref{figR012}
are divided into the four parts, 
each of which have their own physical meaning.
 The neutrino events in the first sector where 
${\cos\theta_{\nu}}>0$ and ${\cos\theta_{\mu}}>0$
consist of the upward neutrinos and upward muons.
  Namely, this denotes such a situation that the 
upward neutrinos emit the muons in forward directions.
 The neutrino events in the second sector where 
${\cos\theta_{\nu}}<0$ and ${\cos\theta_{\mu}}>0$
consist of the downward neutrinos and the upward muons. 
Namely, this denotes such a situation that downward neutrinos emit 
the muons in backward directions. 
The neutrino events in the third sector where 
${\cos\theta_{\nu}}<0$ and ${\cos\theta_{\mu}}<0$
consist of the downward neutrino and the downward muons.
 Namely, this denotes such a situation that the downward neutrinos 
emit the muons in forward directions. 
The neutrino events in the fourth sector where 
${\cos\theta_{\nu}}>0$ and ${\cos\theta_{\mu}}<0$
consist of the upward neutrinos and the downward muons.
 Namely, this denotes such a situation that the upward neutrinos 
emit the muons in backward directions.

Here, the downward muons are generated in three different manners.
In the first case, they are generated by downward neutrinos.
In the second case, they are generated by upward neutrinos,
 owing to the backscattering in QEL. 
And in the third case, they are generated by  either upward or downward 
horizontal-like neutrinos.
 In the last case, either upward or downward horizontal-like neutrinos 
may produce downward  muons due to the accidental azimuthal effect 
in QEL.(see, Figure 3b).   
 The upward muons are generated in three different manners similarly 
by  both upward and downward neutrinos. 
 
 It is clear from Figure~\ref{figR011} that
events in the first sector are symmetrical to those in the 
third sector, 
while events in the second sector are symmetrical to those in the forth 
sector. 
As we expect the same incident neutrino fluxes in both downward and upward 
directions in the case of no oscillation,
 these symmetries are should be hold, as it must be.
 On the other hand, it is clear from Figure~\ref{figR012}
 that symmetries found in Figure~\ref{figR011} are lost any more.
 This comes from that the upward neutrino flux is smaller than that 
of the downward neutrino in the case of the existence of oscillation.    

In Figure~\ref{figR013}, we give the events which exist in the case 
of no oscillation but disappear due to oscillation 
obtained through the Monte Carlo
operational procedure by the survival probability of a given flavor
( Eq.(1) ).
It is simply obtained by the subtraction of the events in 
Figure~\ref{figR012} from the events in Figure~\ref{figR011}.

We can obtain several interesting results from Figure~\ref{figR013}.
 The first is that we do not find disappeared (rejected) events in 
the downward neutrino events due to oscillation. 
This reason is very simple and clear.
It is due to the extreme low probability of oscillation for downward
neutrino with the statistics considered on the occasion
of the selection of the specified neutrino oscillation
paremeters by Super-Kamiokande Collaboration,
say $\Delta m^2 = 2.4\times 10^{-3}\rm{ eV^2}$.
When we adopt this value one order of magnitude larger than SK's,
$\Delta m^2 = 2.4\times 10^{-2}\rm{ eV^2}$,
we surely expect the events concerned for the downward neutrinos.
Furthermore,
when we adopt this value one order of magnitude smaller than SK's,
$\Delta m^2 = 2.4\times 10^{-4}\rm{ eV^2}$,
then we surely expect different distribution of the events concerned
even for the upward neutrinos,
keeping the absence of the events concerned for downward neutrinos.
The second is that almost disappeared (rejected) events belong to the 
first sector (upward neutrinos $\rightarrow$ upward muons). 
The third is that we cannot neglect disappeared (rejected) neutrino 
events which belong to the forth sector 
(upward neutrino $\rightarrow$ downward muon).

\begin{figure}
\begin{center}
\vspace{-1cm}
\hspace*{-2.5cm}
\rotatebox{90}{%
\resizebox{0.5\textwidth}{!}{%
  \includegraphics{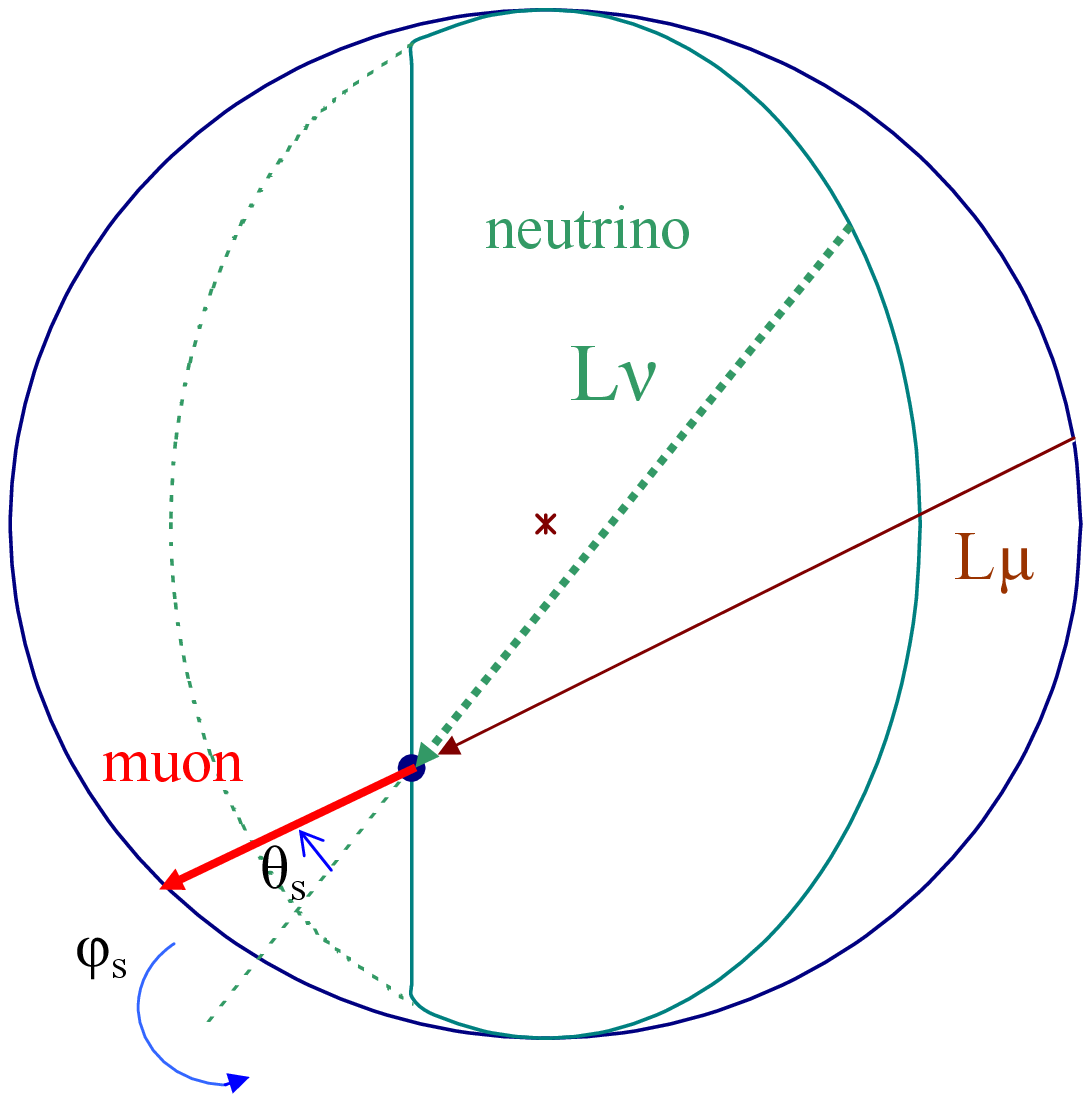}
}}
\vspace{-2cm}
\caption{Schematic view of relations among
$L_{\nu}$, $L_{\mu}$, $\theta_s$ and $\phi_s$
.}
\label{figH010}
\resizebox{0.5\textwidth}{!}{%
  \includegraphics{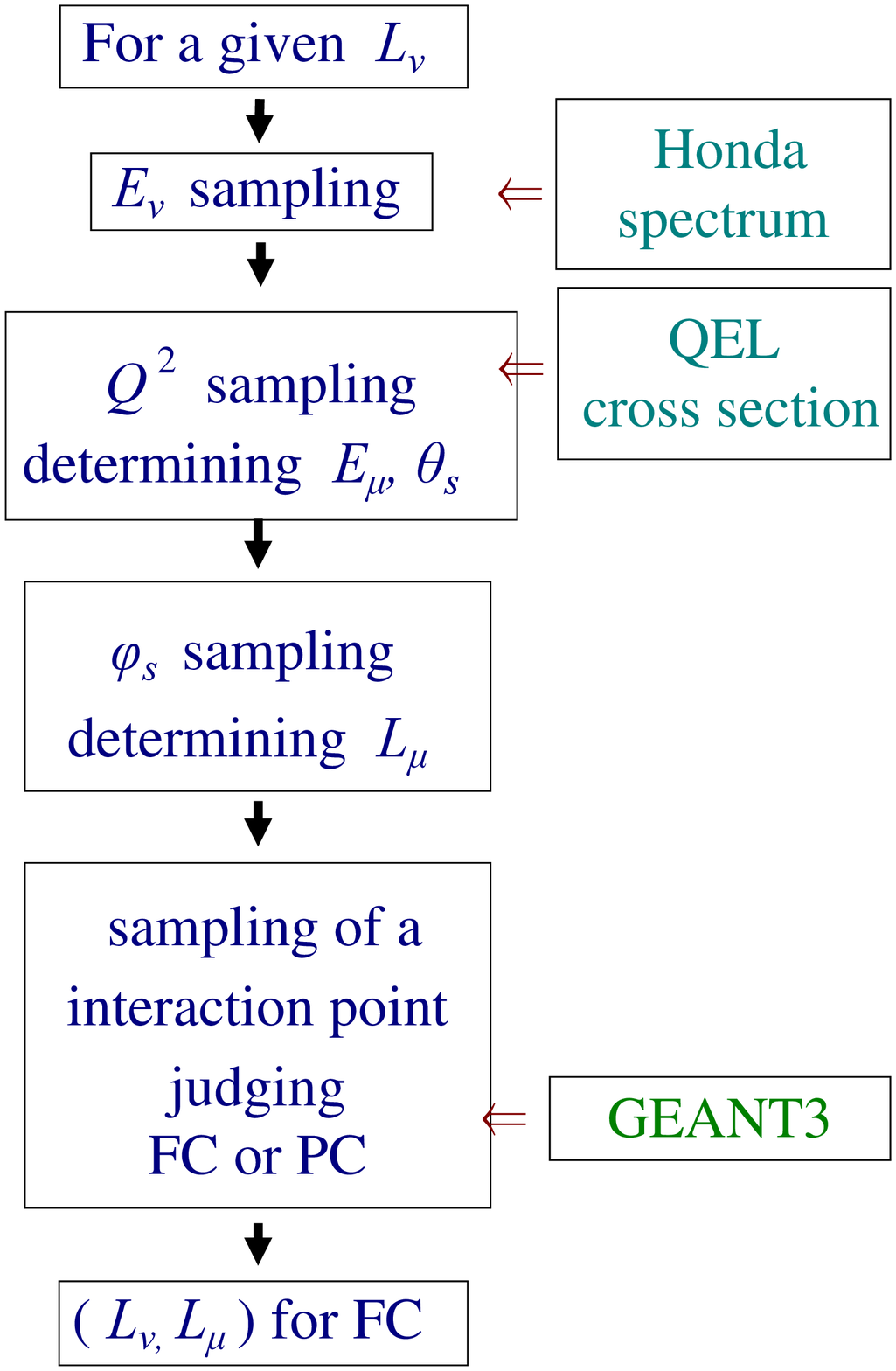}
}
\vspace{-1.5cm}
\caption{The procedure for our numerical experiment
for obtaining $L_{\mu}$ from a given $L_{\nu}$.
}
\label{figH011}
\end{center}
\end{figure}

\subsection{Correlation between 
$L_{\nu}$ and $L_{\mu}$
}

In the previous subsection, we verify that 
{\it the SK assumption on the direction} 
does not hold from the examination on the correlatin between
${\cos\theta_{\nu}}$ and ${\cos\theta_{\mu}}$.

  This SK assumption on ${\cos\theta_{\nu}}$ and ${\cos\theta_{\mu}}$
is logically equivalent to the statement 
that $L_{\nu}$ is approximately the same as $L_{\mu}$ in $L/E$ analysis, 
where $L_{\nu}$ denotes the distance on the incident neutrino from 
the interaction point of the neutrino events to the intersection of 
the Earth surface toward its arriving direction and 
$L_{\mu}$ denotes the corresponding distance on the emitted muon.
Consequently, if our indication on the invalidity of
{\it the SK assumption on the direction} on
${\cos\theta_{\nu}}$ and ${\cos\theta_{\mu}}$ is correct,
 the same conclusion should be drawn from the correlation 
between $L_{\nu}$ and $L_{\mu}$.    
 In the present subsection, we directly examine 
the validity of the implicit SK assumption that $L_{\nu}$ is 
approximated by $L_{\mu}$, 
taking consideration of the neutrino energy spectrum at the
Super-Kamiokande site.
The relation between $L_{\nu}$ and $L_{\mu}$ is given in 
Figure~\ref{figH010}. Also, we show the procedure to determine
 $L_{\mu}$ for a given $L_{\nu}$ in Figure~\ref{figH011}.
\begin{figure}
\begin{center}
\vspace{-1.5cm}
\hspace*{-1.2cm}
\rotatebox{90}{%
\resizebox{0.45\textwidth}{!}{%
  \includegraphics{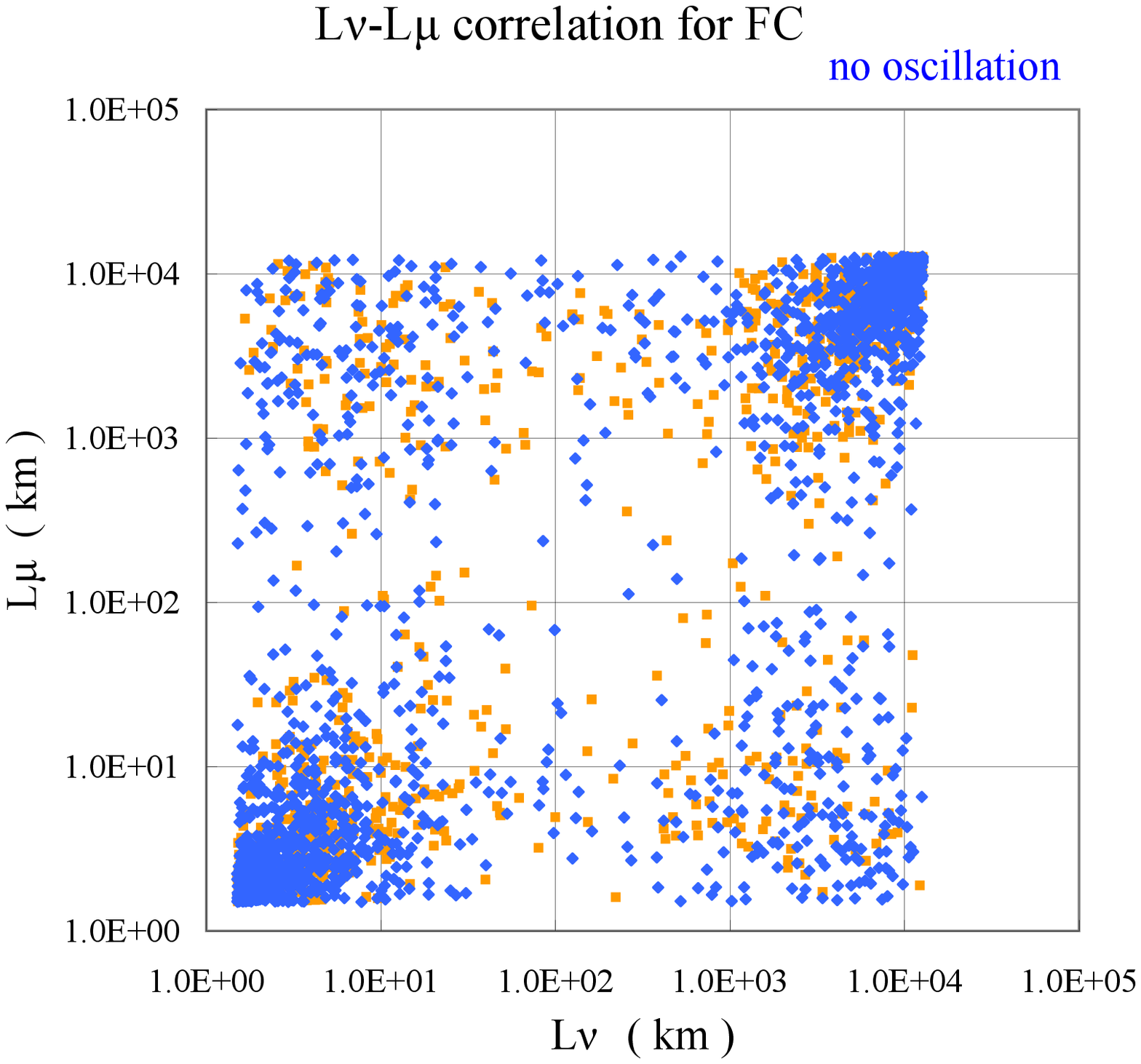}
}}
\vspace{-1.5cm}
\caption{Correlation diagram for $L_{\nu}$ and $L_{\mu}$ without 
oscillation for 1489.2 live days.
The blue points and orange points denote neutrino events and 
ani-neutrino events, respectively.  
}
\label{figH012}       
\vspace{-0.5cm}
\hspace*{-1.2cm}
\rotatebox{90}{%
\resizebox{0.45\textwidth}{!}{%
  \includegraphics{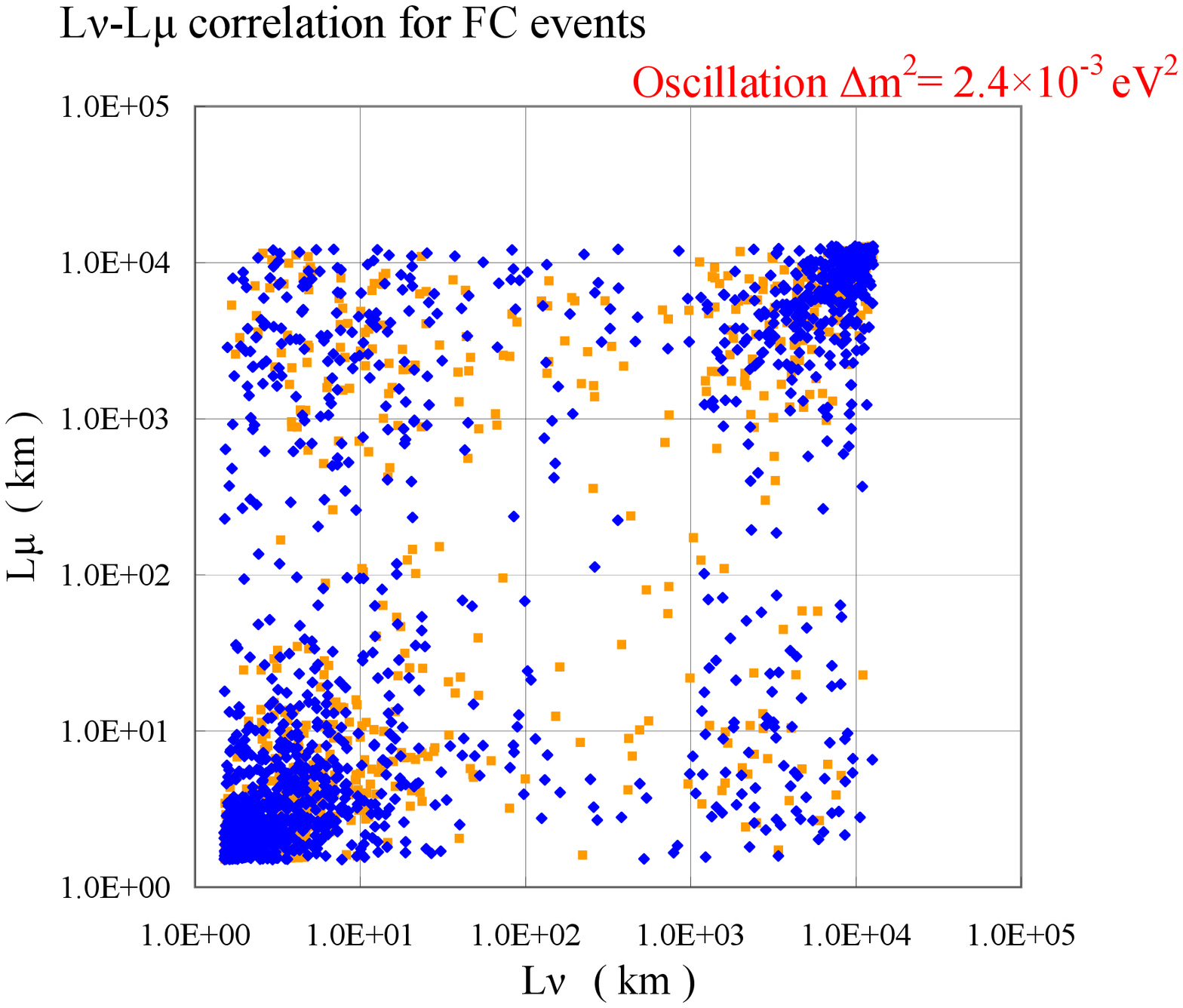}
}}
\vspace{-1.5cm}
\caption{Correlation diagram for $L_{\nu}$ and $L_{\mu}$ with 
 oscillation for 1489.2 live days.
The blue points and orange points denote neutrino events and 
ani-neutrino events, respectively.  
}
\label{figH013}
\vspace{-0.5cm}
\hspace*{-1.2cm}
\rotatebox{90}{%
\resizebox{0.45\textwidth}{!}{%
  \includegraphics{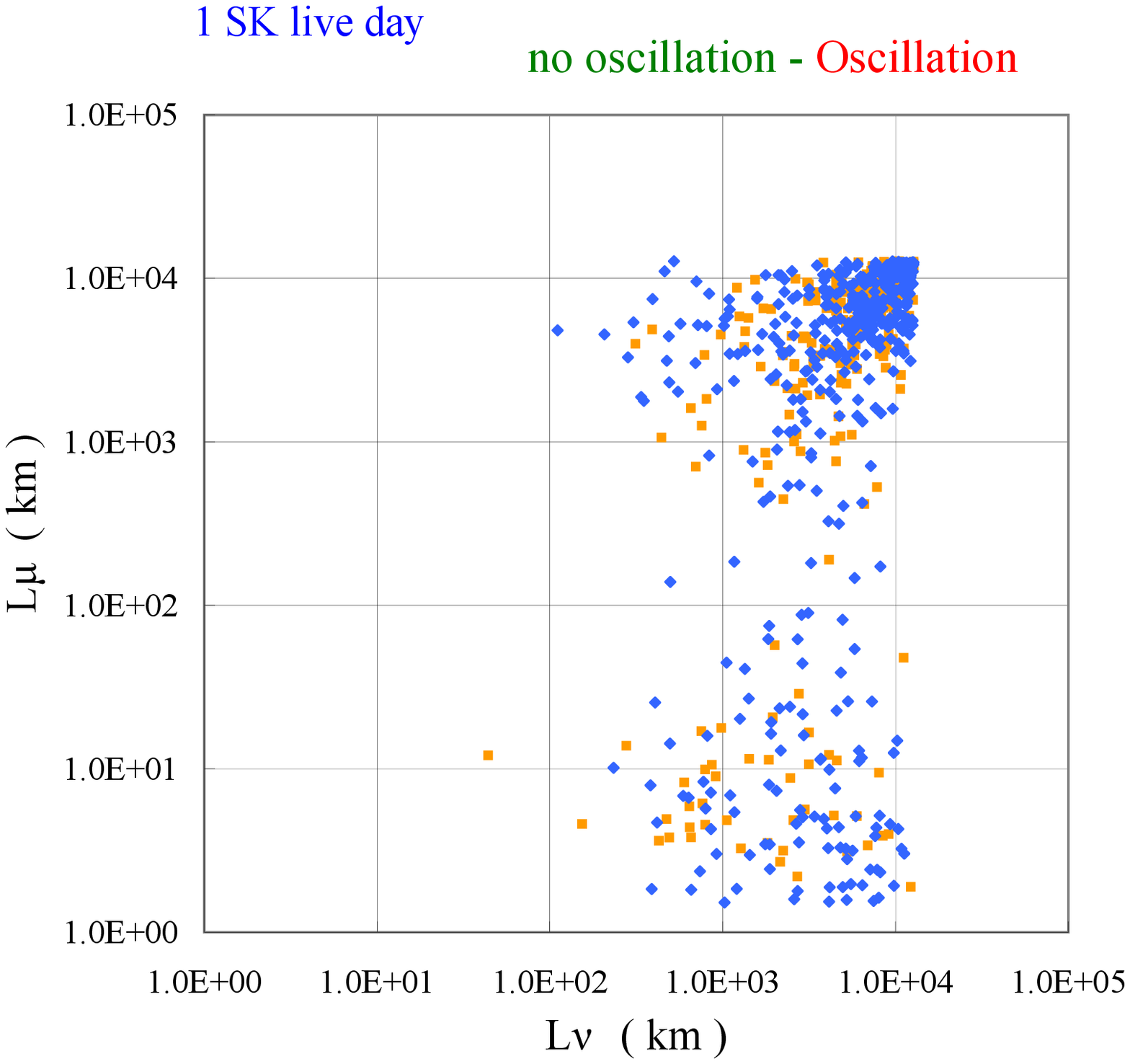}
}}
\vspace{-1.5cm}
\caption{Correlation diagram for $L_{\nu}$ and $L_{\mu}$  
 for the events exist in null oscillation
but disappear due to oscillation 
for 1489.2 live days.
The blue points and orange points denote neutrino events and 
ani-neutrino events, respectively.  
}
\label{figR015}
\end{center}
\end{figure}

$L_{\nu}$ and $L_{\mu}$ are functions of the direction cosine of the 
incident neutrino, $cos\theta_{\nu}$, and that of emitted muon,
 $cos\theta_{\mu}$, respectively and they are given as,
$$
L_{\nu}= R_g \times ( r_{SK} cos\theta_{\nu} +
\sqrt{ r_{SK}^2 cos^2\theta_{\nu} + 1 - r_{SK}^2} )  \,\,\,\,(6-1)
$$
$$
L_{\mu}= R_g \times ( r_{SK} cos\theta_{\mu} +
\sqrt{ r_{SK}^2 cos^2\theta_{\mu} + 1 - r_{SK}^2} )  \,\,\,\,(6-2)
$$
\noindent where $R_g$ is the radius of the Earth and 
$r_{SK}=1-D_{SK}/R_g$, with the depth, $D_{SK}$, 
of the Super-Kamiokande
Experiment detector from the surface of the Earth.
It should be noticed that the $L_{\nu}$ and $L_{\mu}$ are regulated by
both the 
energy spectrum of the incident neutrino and the production spectrum of 
the muon (QEL in the present case). Consequently,
 their mutual relation is influenced by either 
the absence of oscillation or the presence of oscillation which 
depend on the combination of oscillation parameters.



 In Figures~\ref{figH012} and \ref{figH013},
 we give the correlation diagram between
$L_{\nu}$ and $L_{\mu}$ for single ring muon events
among {\it Fully Contained Events} for the 1489.2 live days 
in both the absence and the present of neutrino oscillation
which corresponds to the actual Super-Kamiokande Experiment\cite{Ashie2}.  
In Figure~\ref{figR015}, we give the subtraction of
Figure~\ref{figH013} from Figure~\ref{figH012}
which corresponds to Figure~\ref{figR013}, exactly.
In Figures~\ref{figH012} to \ref{figR015},
blue points denote neutrino events while 
orange points denote anti-neutrino events.
Throughout all correlation diagrams in the present paper,
blue points and orange ones have the same meaning as shown in 
these Figures.
The aggregates of the (anti-) neutrino events which 
correspond to definite 
combinations of $L_{\nu}$ and $L_{\mu}$ are essentially classified into 
four groups  
in the following,
and they correspond to four sections with regards to 
${\cos\theta_{\nu}}$ and ${\cos\theta_{\mu}}$
as shown in Figures~\ref{figR011} to \ref{figR013}:  

 Group A is defined as the aggregate for neutrino events in which 
both $L_{\nu}$ and $L_{\mu}$ are rather small. 
It denotes that the downward neutrinos produce the downward muons
with smaller scattering angles.
 In this case, energies of the produced muons are near those 
 of the incident neutrinos due to smaller scattering angles.

  Group B is defined as the aggregate for neutrino events in which 
both $L_{\nu}$ and $L_{\mu}$ are rather large.
It denotes that the upward neutrinos  produce upward muons
with smaller scattering angles.
In this case, the energy relation between incident neutrinos and 
 produced muons is essentially the same as in Group A 
in Figure~\ref{figH012}, because the
flux of the upward neutrino events is symmetrical to that of 
downward neutrino events in the absence of neutrino oscillation.

 Group C is defined as the aggregate for neutrino events in 
which $L_{\nu}$ are rather small and $L_{\mu}$ are rather large.
 It denotes that the downward neutrinos produce the 
upward muons by the possible effect resulting from
 both backscattering and azimuthal angle in QEL.
 In this case, energies of the produced muons are smaller 
than those of the incident neutrinos 
due to larger scattering angles.

 Group D is defined as the aggregate for the neutrino events in 
which $L_{\nu}$ are rather large and $L_{\mu}$ are rather small.
 It denotes that the upward neutrinos produce the 
downward muons. The energy relation between the incident neutrinos and 
the produced muons is essentially the same as in Group C in the absence 
of neutrino oscillation (Figure~\ref{figH012}).

Summarized from the mentioned above, we can say
 that there exist the 
symmetries  between Group A and Group B, and also between Group C and 
Group D, which reflect  
the symmetry between the upward neutrino flux and the downward neutrino 
one for null oscillation.  

 In Figure~\ref{figH013}, we give the correlation between 
$L_{\nu}$ and $L_{\mu}$
under their neutrino oscillation parameters, say,
$\Delta m^2 = 2.4\times 10^{-3}\rm{ eV^2}$ and $sin^2 2\theta=1.0$
\cite{Ashie2}. 
In the presence of neutrino oscillation, the property of the 
symmetry which holds in the absence of neutrino oscillation 
(see $\langle$Group A and Group B$\rangle$ and/or 
$\langle$Group C and Group D$\rangle$ in Figure~\ref{figH012})
is lost due to different incident neutrino fluxes in the upward 
direction and the downward one. 
If we compare Group A with Group B, the event 
number in Group B (upward $\nu$ $\rightarrow$ upward $\mu$)
is smaller than that in group A
(downward $\nu$ $\rightarrow$ downward $\mu$),
 which comes from smaller flux of the upward neutrinos. 
The similar relation between Group C 
(downward $\nu$ $\rightarrow$  upward $\mu$ ) and Group D 
(upward $\nu$ $\rightarrow$  downward $\mu$) holds in 
Figure~\ref{figH013}.

 Summarizing the characteristics among Groups A to D in the 
Figures~\ref{figH012} and \ref{figH013}, we can conclude that 
$\langle$Group A and Group B$\rangle$ and 
$\langle$Group C and Group D$\rangle$ are in symmetrical situations 
in the absence of neutrino oscillation,
while such a symmetrical situation is lost in the presence 
of neutrino oscillation.
  Also, it is clear from  Figures~\ref{figH012} 
and \ref{figH013} that $L_{\nu}\approx L_{\mu}$, namely  
{\it the SK assumption on the direction}, does not hold 
both in the absence of neutrino oscillation and 
in the presence of neutrino oscillation
even if statistically.

Here, it should be noticed that the approximation of
$L_{\nu}\approx L_{\mu}$ does not hold completely 
in the regions C and D.
The event numbers in Group C and Group D could not be neglected 
among the total event number concerned. 
 In these regions, neutrino events consist of those with 
backscattering and/or neutrino events 
in which the neutrino directions
are horizontally downward (upward),
 but their produced muons turn upward (downward) 
resulting from the effect of azimuthal angles in QEL. 

In the $L/E$ analysis made by Super-Kamiokande Collaboration,
 the reconstruction of the direction of the incident neutrino 
from the direction of the emitted muon is very simple, say,
\[
\hspace{2cm}\cos\Theta^{rec}_{\nu}=\cos\Theta_{\mu} 
\]
and there is nothing more
\footnote{
following 
"$cos\Theta^{rec}_{\nu}=\cos\Theta_{\mu}$ (8.17)" ,
Ishitsuka states "$cos\Theta^{rec}_{\nu}$ and $\cos\Theta_{\mu}$
are cosines of the reconstructed zenith angles of neutrino and muon,
respectively." (see text in page 3)
.}.

\begin{figure*}
\vspace{-3.5cm}
\hspace*{-1cm}
\rotatebox{90}{%
\resizebox{0.4\textwidth}{!}{%
  \includegraphics{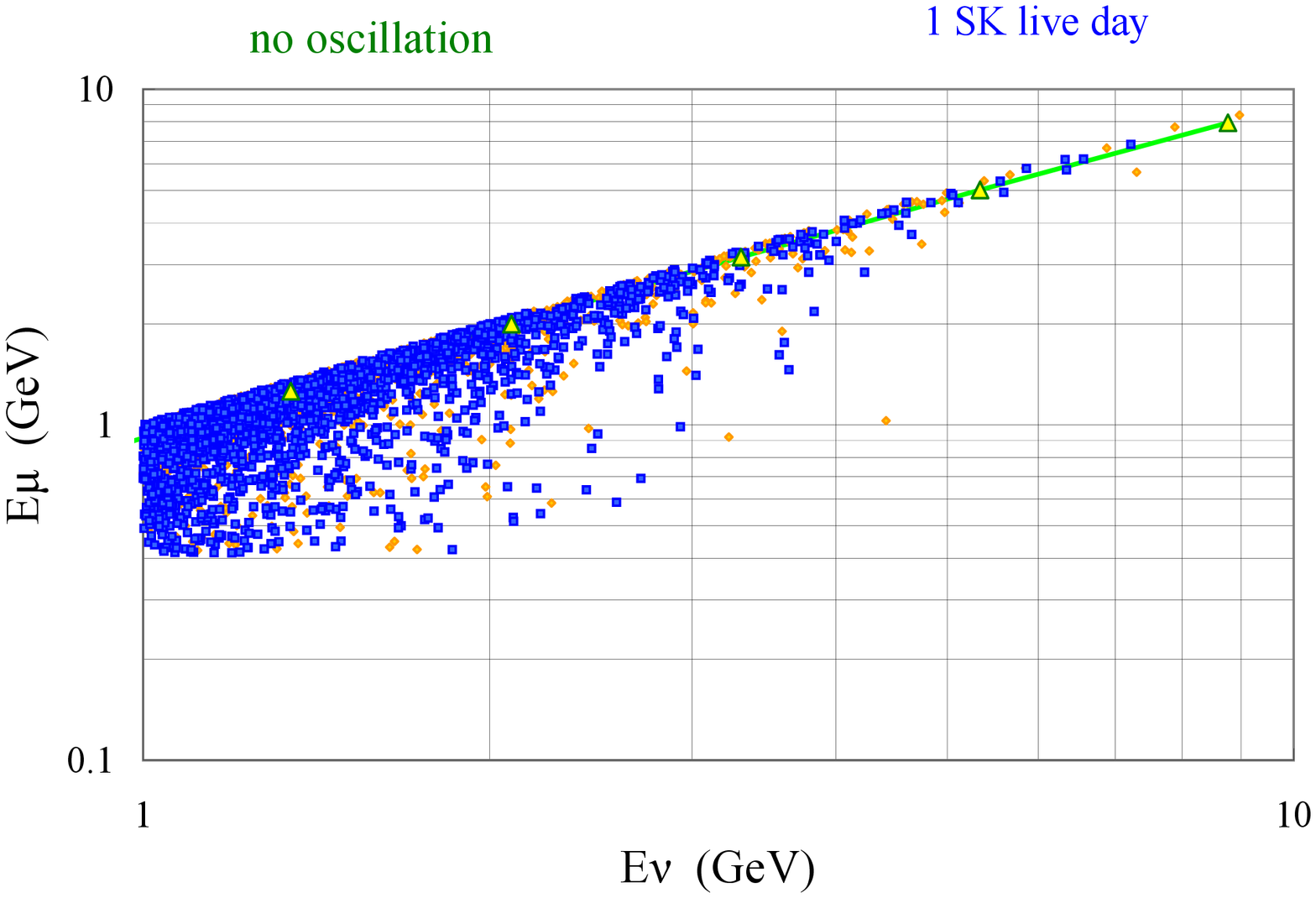}
}}
\hspace*{-1cm}
\rotatebox{90}{%
\resizebox{0.4\textwidth}{!}{%
  \includegraphics{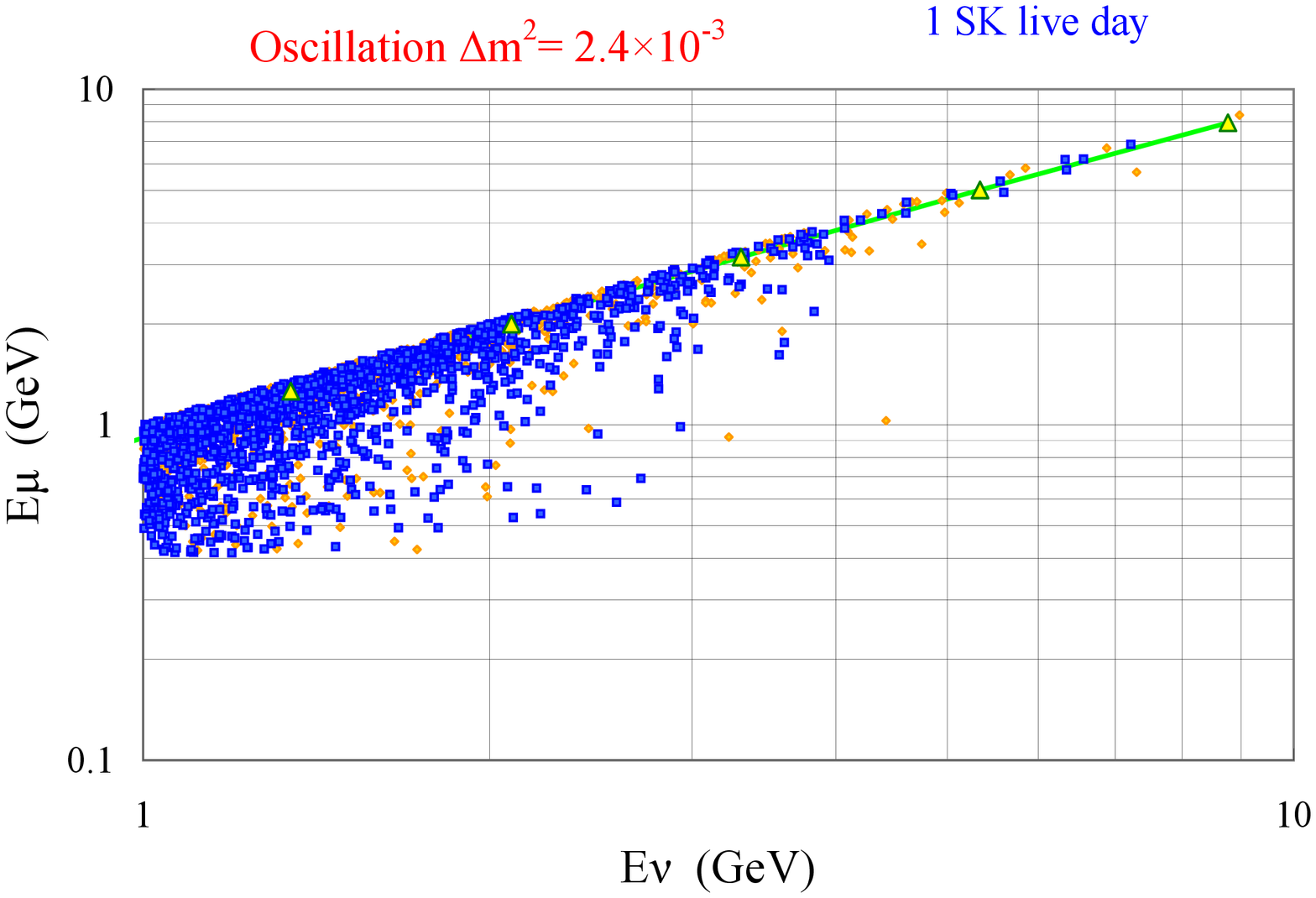}
}}
\vspace*{-1.5cm}
\begin{center}
\hspace*{-1cm}(a)
\hspace*{8cm}(b)
\vspace*{-1.5cm} \\
\rotatebox{90}{%
\resizebox{0.4\textwidth}{!}{%
  \includegraphics{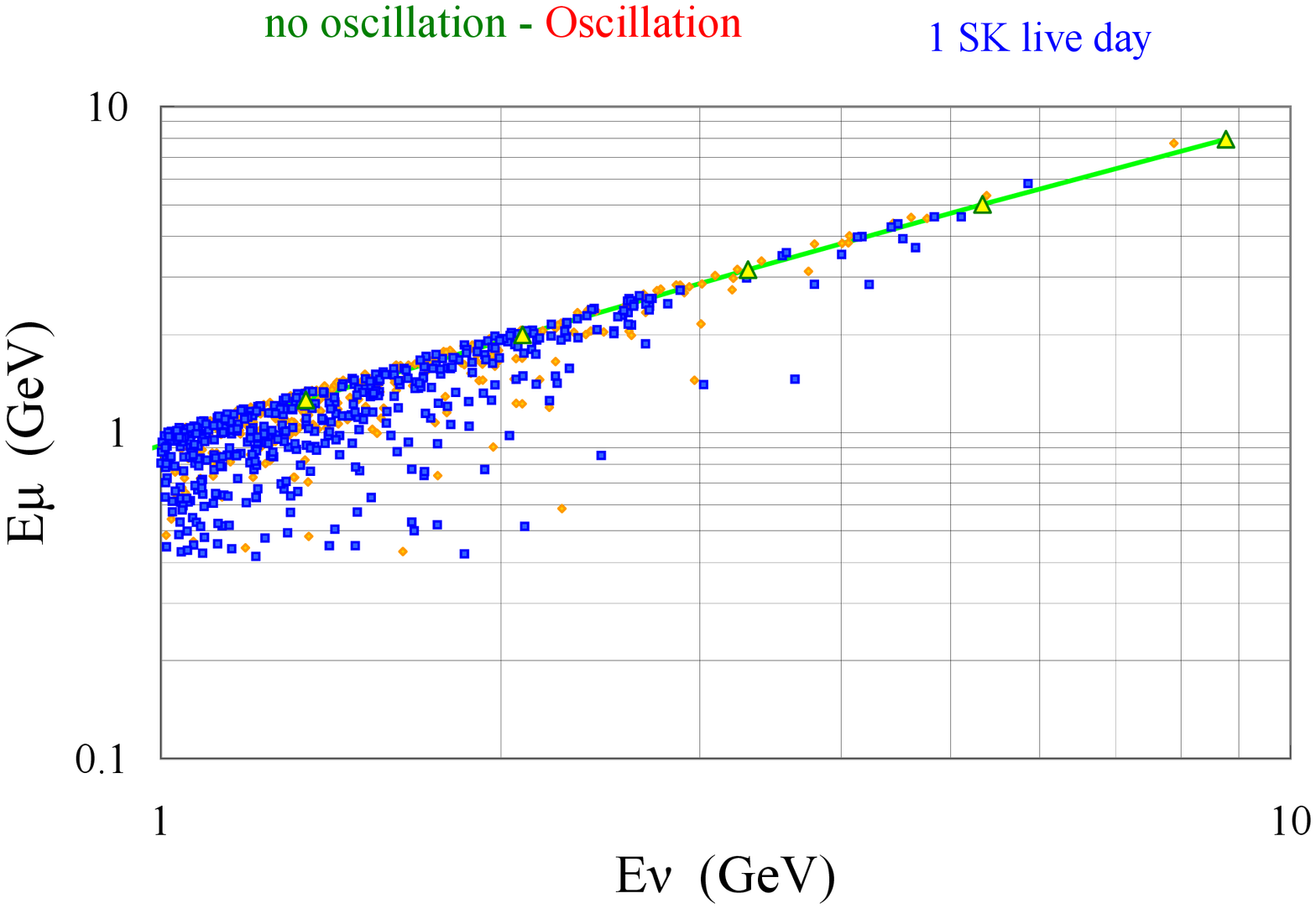}
}}\\
\vspace*{-0.7cm}
(c)
\caption{
The correlation diagram 
between $E_{\nu}$ and $E_{\mu}$ (a) without 
oscillation, (b) with oscillation and
(c) for the events which exist in null oscillation
but disappear due to oscillation,  
for 1489.2 live days, respectively.
  The solid line denotes 
the polynomial expression by Super-Kamiokande Collabolation.
}
\label{figH014}
\end{center}
\end{figure*}

 Of course, as we pointed out already,
Super-Kamiokande Collaboration know the existence of 
the non-negligible scattering angles of the neutrino interaction.
 However, they don't consider it in their analysis.  
The concept of the "reconstruction" by Super-Kamiokande Collaboration 
denote one-to-one correspondence between the direction of the 
incident neutrino and the direction of the emitted lepton.
 However, the correlations between them shown in 
Figures~\ref{figH012} and \ref{figH013}
make it impossible to reconstruct the direction of the incident 
neutrino from the direction of emitted muon,
even if one takes the scattering angle into account, 
because these correlations are the reflection of the stochastic 
characters in the processes concerned which deny unique relation 
between them.
Again, we should emphasize and confirm that
 "reconstruction of the direction of the incident neutrino"
 in $L/E$ analysis of Super-Kamiokande Collaboration 
 is just equal to the adoption of the direction of the emitted muon 
due to the neutrino interaction.

\subsection{The correlation between $E_{\nu}$ and $E_{\mu}$}

Here, we examine the transformation of $E_{\nu}$ from $E_{\mu}$
adopted by Super-Kamiokande Collaboration.
 The validity problem is closely related to 
{\it the SK assumption on the direction}
 through the survival probability for a given flavor.

 Super-Kamiokande Collaboration estimate $E_{\nu}$ from $E_{\mu}$,
 the visible energy of the muon,
 from their Monte Carlo simulation,
by the following equation\cite{Ishitsuka}
(see page 135 of the paper concerned) : 
$$
E_{\nu,SK}= E_{\mu}\times(a + b\times x + c\times x^2 + 
d\times x^3 ),  \,\,\,(7)
$$

\noindent where $x = log_{10}(E_{\mu})$.

 The idea that $E_{\nu}$ can be approximated as the polynomial of
$E_{\mu}$ means 
that there should be the unique relation between $E_{\nu}$ and $E_{\mu}$.
 However, in the light of stochastic characters inherent in both the 
incident neutrino energy spectrum and the production spectrum of 
the muon, such a treatment is not suitable theoretically,
 which may kill a real correlation effect between the incident 
neutrino energy and the emitted muon energy.
 In Figures~\ref{figH014}, we give the correlation between 
$E_{\nu}$ and $E_{\mu}$ 
together with that obtained from the polynomial expression by 
Super-Kamiokande Collaboration under 
their neutrino oscillation parameters 
and their incident neutrino energy spectrum\cite{Honda}. 
It is clear from the figure that the part of the lower energy 
incident neutrino deviates largely from the approximated formula,
 which reflects explicitly the stochastic character of QEL.

In Figure~\ref{figH014}~(b), we give the correlation between
$E_{\nu}$ and $E_{\mu}$ with oscillation.
Glancing at Figures~\ref{figH014}~(a) and (b),
we cannot recognize the difference between them,
because too much events are marked on the figure for their discrimination.
Then in Figure~\ref{figH014}~(c), we give the subtraction between them,
as shown in Figures~\ref{figR013} and \ref{figR015}.
It is easily understood that 
there is the visible difference between them.
Furthermore, it is easily understood from Figures~\ref{figH014} that
the polynomial expression is not suitable for the description on the
mutual relation between $E_{\nu}$ and $E_{\mu}$, standing on the
stochastic point of view. 

Finally, it is necessary to mention to the qualitative difference 
between the correlation between $L_{\nu}$ and $L_{\mu}$, and the 
correlation between $E_{\nu}$ and $E_{\mu}$,
 from the degree of their influence on 
{\it the SK assumption on the direction}.
  As clarified in  
Figures~\ref{figH012} and \ref{figH013},
it is impossible to approximate 
as $L_{\nu}$ nearly equal to $L_{\mu}$ ($L_{\nu} \approx L_{\mu}$). 
On the other hand, the approximation of $E_{\nu}$ by Eq.(7) does not 
introduce fatal error finally, 
although its treatment is not theoretically suitable in the sense 
of the lack of stochastic characters on the physical processes.
 Namely, what influences on the $L/E$ analysis essentially is 
the correlation between $L_{\nu}$ and $L_{\mu}$, 
but not the correlation between $E_{\nu}$ and $E_{\mu}$.   


\section{Conclusion}
Since one cannot measure $L_{\nu}$ and $E_{\nu}$, one is forced to utilize
$L_{\mu}$ and $E_{\mu}$  
in the $L/E$ analysis in place of them.
 Then,  
Super-Kamiokande Collaboration 
assume that the direction of the incident neutrino is 
approximately the same as 
that of the emitted lepton, namely 
{\it the SK assumption on the direction} and
$E_{\nu}$ can be estimated from some polynomial formula of the 
variable $E_{\mu}$ in their $L/E$ analysis.
However, it is clear from 
Figures~\ref{figR012} and \ref{figR013},
and/or Figures~\ref{figH012} and \ref{figH013},
that {\it the SK assumption on the direction}
 does not hold even approximately and the transformation of
$E_{\mu}$ into $E_{\nu}$ is not uniquely.

Consequently, we can conclude that $L_{\nu}$ cannot be 
reconstructed by $L_{\mu}$,
 but  $E_{\nu}$ can be done if we allow considerable uncertainties. 

In the subsequent paper (Part 2), we apply the results from 
Figures~\ref{figH012} to \ref{figH014} to $L/E$ analysis and 
conclude that one cannot
 obtain the maximum oscillation in $L/E$ analysis 
reported by Super-Kamiokande which shows
 strongly the oscillation pattern due to neutrino oscillation.

\newpage
\noindent {\bf APPENDICES}\\
\appendix
\section{ 
 Monte Carlo Procedure for the Decision of Emitted Energies of the Leptons and Their Direction Cosines }
\setcounter {equation} {0}
\setcounter {figure} {0}
\def\theequation{\Alph{section}\textperiodcentered\arabic{equation}}

Here, we give the Monte Carlo Simulation procedure for obtaining the energy and its  direction cosines,
$(l_{r},m_{r},n_{r})$, of the emitted lepton in QEL for a given energy and its direction cosines, $(l,m,n)$, of the incident neutrino. 

\begin{figure}
\begin{center}
\resizebox{0.45\textwidth}{!}{%
  \includegraphics{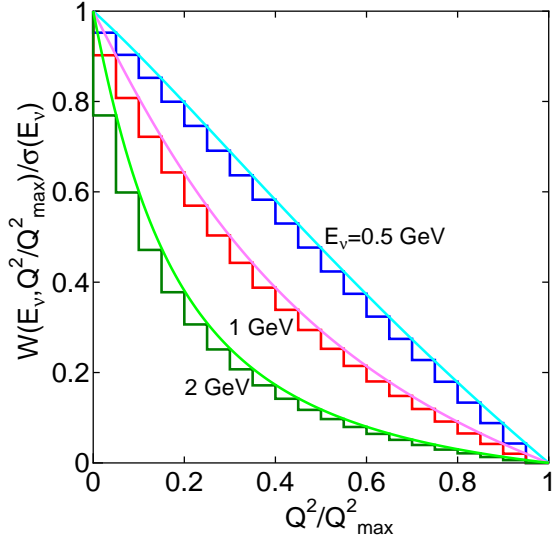}
  }
\end{center}
\caption{ The reappearance of the probability function for QEL cross section.
Histograms are  sampling results, while the curves  concerned are theoretical
ones for given incident energies.
}
\label{figP001}
\end{figure}

The relation among $Q^2$, $E_{\nu(\bar{\nu})}$, the energy of the incident neutrino, $E_{\ell(\bar{\ell})}$, the energy of the emitted lepton (muon or electron or their anti-particles) and $\theta_{\rm s}$, the scattering angle of the emitted lepton, is given as
      \begin{equation}
         Q^2 = 2E_{\nu(\bar{\nu})}E_{\ell(\bar{\ell})}(1-{\rm cos}\theta_{\rm s}).
\label{eqn:a1}  
      \end{equation}
\noindent Also, the energy of the emitted lepton is given by
      \begin{equation}
         E_{\ell(\bar{\ell})} = E_{\nu(\bar{\nu})} - \frac{Q^2}{2M}.
\label{eqn:a2}  
      \end{equation}
\noindent {\bf Procedure 1}\\
\noindent
We decide  $Q^2$ from the probability function for the differential cross section with a given $E_{\nu(\bar{\nu})}$ (Eq. (\ref{eqn:2}) in the text) by using the uniform random number, ${\xi}$,  between (0,1) in the following\\
  \begin{equation}
    \xi = \int_{Q_{\rm min}^2}^{Q^2}P_{\ell(\bar{\ell})}(E_{\nu(\bar{\nu})},Q^2)
                             {\rm d}Q^2,
\label{eqn:a3}  
  \end{equation}
\noindent where
  \begin{eqnarray}
\lefteqn{     P_{\ell(\bar{\ell})}(E_{\nu(\bar{\nu})},Q^2) =} \nonumber \\
&&  \frac{ {\rm d}\sigma_{\ell(\bar{\ell})}(E_{\nu(\bar{\nu})},Q^2) }{{\rm d}Q^2} 
                     \Bigg /\!\!\!\!
      \int_{Q_{\rm min}^2}^{Q_{\rm max}^2} 
      \frac{ {\rm d}\sigma_{\ell(\bar{\ell})}(E_{\nu(\bar{\nu})},Q^2) }{{\rm d}Q^2} 
             {\rm d}Q^2 . \nonumber \\
&&
\label{eqn:a4}  
   \end{eqnarray}
\\
\noindent From Eq. (A$\cdot$1), we obtain $Q^2$ in histograms together with the corresponding theoretical curve shown in Figure~\ref{figP001}. The agreement between the sampling data and the theoretical curve is excellent, which shows the validity of the utlized  procedure in Eq. (A$\cdot$3). \\

\begin{figure}
\begin{center}
\vspace{-1cm}
\resizebox{0.45\textwidth}{!}{%
  \includegraphics{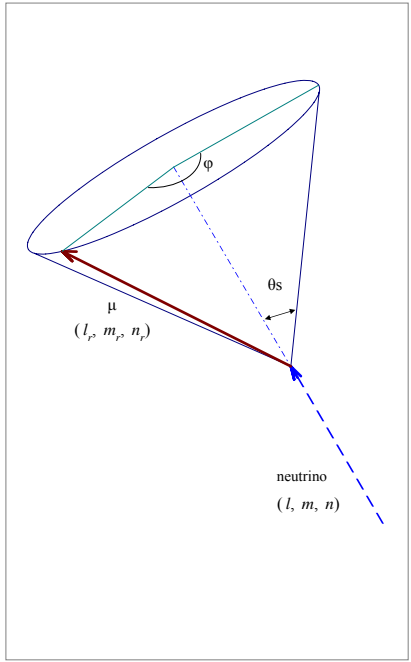}
  }
\end{center}
\caption{ The relation between the direction cosine of the incident neutrino and that of the emitted charged lepton.}
\label{figP002}
\end{figure}

\noindent {\bf Procedure 2}\\
\noindent
We obtain $E_{\ell(\bar{\ell})}$ from Eq. (A$\cdot$2) for  the given $E_{\nu(\bar{\nu})}$ and $Q^2$ thus decided in Procedure 1.\\

\noindent {\bf Procedure 3}\\
\noindent
We obtain $\cos{\theta_{\rm s}}$, cosine of the the scattering angle of the emitted lepton, for $E_{\ell(\bar{\ell})}$ thus decided in the Procedure 2 from Eq. (A$\cdot$1) .\\

\noindent {\bf Procedure 4}\\
\noindent
We decide $\phi$, the azimuthal angle of the scattered lepton, which is obtained from\\
  \begin{equation}
       \phi = 2\pi\xi.                 
\label{eqn:a5}  
  \end{equation}
\noindent Here, $\xi$ is a uniform random number between (0, 1). \\
As explained schematically in the text(see Figure~{\bf \ref{figH003}} 
in the text),  we must take account of the effect due to the azimuthal 
angle $\phi$ in QEL to obtain the zenith angle distribution both
for {\it Fully Contained Events} and {\it Partially Contained Events} 
correctly.\\  

\noindent {\bf Procedure 5}\\
\noindent
The relation between direction cosines of the incident neutrino, $(\ell_{\nu(\bar{\nu})}, m_{\nu(\bar{\nu})}, n_{\nu(\bar{\nu})} )$, and those of the corresponding emitted lepton, $(\ell_{\rm r}, m_{\rm r}, n_{\rm r})$, for a certain $\theta_{\rm s}$ and $\phi$ is given as
\begin{eqnarray}
\lefteqn{
\left(
         \begin{array}{c}
             \ell_{\rm r} \\
             m_{\rm r} \\
             n_{\rm r}
         \end{array}
       \right)
           =
}
\nonumber\\
    &&   \left(
         \begin{array}{ccc}
           \displaystyle \frac{\ell n}{\sqrt{\ell^2+m^2}} & 
            -\displaystyle 
            \frac{m}{\sqrt{\ell^2+m^2}}        & \ell_{\nu(\bar{\nu})} \\
            \displaystyle \frac{mn}{\sqrt{\ell^2+m^2}} & \displaystyle 
            \frac{\ell}{\sqrt{\ell^2+m^2}}     & m_{\nu(\bar{\nu})}    \\
                        -\sqrt{\ell^2+m^2} & 0 & n_{\nu(\bar{\nu})}
         \end{array}
       \right) \times
\nonumber\\       
 &&      \times
\left(
          \begin{array}{c}
            {\rm sin}\theta_{\rm s}{\rm cos}\phi \\
            {\rm sin}\theta_{\rm s}{\rm sin}\phi \\
            {\rm cos}\theta_{\rm s}
          \end{array}
       \right),
\label{eqn:a6}
\end{eqnarray}

\noindent where $n_{\nu(\bar{\nu})}={\rm cos}\theta_{\nu(\bar{\nu})}$,
 and $n_{\rm r}={\rm cos}\theta_{\ell}$. 
Here, $\theta_{\ell}$ is the zenith angle of the emitted lepton. \\

The Monte Carlo procedure for the determination of $\theta_{\ell}$ of the emitted lepton for the parent (anti-)neutrino with given $\theta_{\nu(\bar{\nu})}$ and $E_{\nu(\bar{\nu})}$ involves the following steps:\\

We obtain $(\ell_r, m_r, n_r)$ by using Eq. (\ref{eqn:a6}). The $n_r$ is the cosine of the zenith angle of the emitted lepton which should be contrasted with $n_{\nu}$, that of the incident neutrino.
\\
Repeating the procedures 1 to 5 just mentioned above, we obtain the zenith angle distribution of the emitted leptons for a given zenth angle of the incident neutrino with a definite energy. \\

In the SK analysis,  instead of Eq. (\ref{eqn:a6}), they assume 
$n_r = n_{\nu(\bar{\nu})} $ 
uniquely for ${E_{\mu(\bar{\mu})}} \geq$ 400 MeV.

\section{ Monte Carlo Procedure to Obtain the Zenith Angle of the Emitted Lepton for a Given Zentith Angle of the Incident Neutrino}

  The present simulation procedure for a given zenith angle of the incident neutrino starts from the atmospheric neutrino spectrum at the opposite site of the Earth to the SK detector. We define,
 $N_{\rm int,no-osc}(E_{\nu(\bar{\nu})},t,{\rm cos}\theta_{\nu(\bar{\nu})})$,
 the interaction neutrino spectrum at the depth $t$ from the SK detector 
for the case no oscillation
in the following way,
   \begin{eqnarray}
    \lefteqn{  N_{\rm int,no_-osc}(E_{\nu(\bar{\nu})},t,
{\cos}\theta_{\nu(\bar{\nu})}) =}\nonumber \\
&&N_{\rm sp}(E_{\nu(\bar{\nu})},
\cos\theta_{\nu(\bar{\nu})}) \times \nonumber \\ 
&&  \Bigg(1-\frac{{\rm d}t}{\lambda_1(E_{\nu(\bar{\nu})},t_1,\rho_1)} \Bigg)  \times\nonumber \\
 &&\times\cdots \times \Bigg(1-\frac{{\rm d}t}{\lambda_n(E_{\nu(\bar{\nu})},t_n,\rho_n)} \Bigg).\nonumber \\
&& 
\label{eqn:b1}
   \end{eqnarray}

Here, $N_{\rm sp}(E_{\nu(\bar{\nu})},\cos\theta_{\nu(\bar{\nu})})$ is the atmospheric (anti-)  neutrino spectrum for the zenith angle at the opposite surface of the Earth.
And $\lambda_i(E_{\nu(\bar{\nu})},t_i,\rho_i)$ 
denotes the mean free path from QEL for the neutrino (anti neutrino) 
with the energy $E_{\nu(\bar{\nu})}$ at the distance, $t_i$, from the opposite surface of the Earth, where $\rho_i$ is there density. 

In the presence of oscillation, neutrino energy spectrum
correponding to (B-1) is given as,

   \begin{eqnarray}
    \lefteqn{  N_{\rm int,osc}(E_{\nu(\bar{\nu})},t,
{\cos}\theta_{\nu(\bar{\nu})}) }\nonumber\\
&&=N_{\rm int,no_-osc}(E_{\nu(\bar{\nu})},
\cos\theta_{\nu(\bar{\nu})}) \times 
P(\nu_{\mu} \rightarrow \nu_{\mu}) 
 \nonumber\\ 
\label{eqn:b2}
   \end{eqnarray}

Here, $P(\nu_{\mu} \rightarrow \nu_{\mu})$ is the survival 
probability of a given flavor, such as $\nu_{\mu}$, and it is given by
                                                    
\begin{eqnarray}
\lefteqn{P(\nu_{\mu} \rightarrow \nu_{\mu})=}
 \nonumber \\
&& 1- sin^2 2\theta \cdot sin^2
(1.27\Delta m^2 L_{\nu} / E_{\nu} ),  
\\ \nonumber 
\end{eqnarray}                                                    

where $sin^2 2\theta = 1.0$ and
$\Delta m^2 = 2.4\times 10^{-3}\rm{eV^2}$ obtained from 
Super-Kamiokande Collaboration\cite{Ashie2}.

The procedures of the Monte Carlo Simulation for the incident neutrino(anti neutrino) with a given energy, $E_{\nu(\bar{\nu})}$, whose incident direction is expressde by $(l,m,n)$ are as follows.\\

\noindent {\bf Procedure A}\\
\noindent
For the given zenith angle of the incident neutrino, 
${\theta_{\nu(\bar{\nu})}}$, we formulate, 
$N_{\rm pro}( E_{\nu(\bar{\nu})},t,\cos\theta_{\nu(\bar{\nu})}){\rm d}E_{\nu(\bar{\nu})}$,
 the production function for the neutrino flux to produce leptons at the 
Kamioka site in the following
   \begin{eqnarray}
\lefteqn{N_{\rm pro}( E_{\nu(\bar{\nu})},t,\cos\theta_{\nu(\bar{\nu})}){\rm d}E_{\nu(\bar{\nu})} } \nonumber \\
&&=
 \sigma_{\ell(\bar{\ell})}(E_{\nu(\bar{\nu})}) N_{\rm int}(E_{\nu(\bar{\nu})},t,{\rm cos}\theta_{\nu(\bar{\nu})}){\rm d}E_{\nu(\bar{\nu})},
\nonumber\\
\label{eqn:b2}
   \end{eqnarray}
\noindent where
  \begin{eqnarray}
     \displaystyle
      \sigma_{\ell(\bar{\ell})}(E_{\nu(\bar{\nu})}) = \int^{Q_{\rm max}^2}_{Q_{\rm min}^2}  \frac{ {\rm d}\sigma_{\ell(\bar{\ell})}(E_{\nu(\bar{\nu})},Q^2)}{{\rm d}Q^2}{\rm d}Q^2.
\label{eqn:b3}
\nonumber\\
  \end{eqnarray}

\noindent \\
Each differential cross section above is given by Eq. (\ref{eqn:2}) in the text.
Here, we simply denote the interaction energy spectrum as
$ N_{\rm int}(E_{\nu(\bar{\nu})},t,
{\cos}\theta_{\nu(\bar{\nu})}) $
, irrespective of the absence or the presence of 
oscillation.

Utilizing, $\xi$, the uniform random number between (0,1), 
we determine $E_{\nu(\bar{\nu})}$, the energy of the incident neutrino 
in the following sampling procedure\\
    \begin{eqnarray}
       \xi = \int_{E_{\nu(\bar{\nu}),{\rm min}}}^{E_{\nu(\bar{\nu})}}
             P_d(E_{\nu(\bar{\nu})},t,\cos\theta_{\nu(\bar{\nu})}(\bar{\nu})){\rm d}E_{\nu(\bar{\nu})},
\nonumber\\
    \end{eqnarray}
where
\begin{eqnarray}
\lefteqn{
        P_d(E_{\nu(\bar{\nu})},t,\cos{\theta}_{\nu(\bar{\nu})}){\rm d}E_{\nu(\bar{\nu})} 
        } \nonumber\\
&=& 
\frac{
N_{pro}( E_{\nu(\bar{\nu})},t,\cos{\theta}_{\nu(\bar{\nu})}){\rm d}E_{\nu(\bar{\nu})} 
}
{ \displaystyle \int_{E_{\nu(\bar{\nu}),{\rm min}}}^{E_{\nu(\bar{\nu}),{\rm max}}} 
                       N_{pro}( E_{\nu(\bar{\nu})},t,\cos{\theta}_{\nu(\bar{\nu})}){\rm d}E_{\nu(\bar{\nu})} 
} .
\nonumber\\
\end{eqnarray}
\\
In our Monte Carlo procedure, the reproduction of, 
$P_d(E_{\nu(\bar{\nu})},t,\cos\theta_{\nu(\bar{\nu})}){\rm d}E_{\nu(\bar{\nu})}$, 
the normalized differential neutrino interaction probability function, is confirmed in the same way as in Eq. (A$\cdot$4). 
\\
%
%
%
%
 
\noindent {\bf Procedure B}\\
\noindent
For the (anti-)neutrino concerned with the energy of $E_{\nu(\bar{\nu})}$, 
we sample $Q^2$ utlizing $\xi$, the uniform random number between (0,1). 
The Procedure B is exactly the same as Procedure 1 in Appendix A. \\

\noindent {\bf Procedure C}\\
\noindent
We decide, ${\theta_{\rm s}}$, the scattering angle of the emitted lepton 
for given $E_{\nu(\bar{\nu})}$ and $Q^2$. Procedure C is exactly the same 
as in the combination of Procedures 2 and 3 in Appendix A. \\

\noindent {\bf Procedure D}\\
\noindent
We randomly sample the azimuthal angle of the charged lepton concerned. 
The Procedure D is exactly the same as in Procedure 4 in Appendix A. \\
 
\noindent {\bf Procedure E}\\
\noindent
We decide the direction cosine of the charged lepton concerned.  
Procedure E is exactly the same as  Procedure 5 in Appendix A.\\

 We repeat Procedures A to E until we reach the desired trial number. \\
\\

\section{Correlation between the Zenith Angles of the Incident Neutrinos and Those of the Emitted Leptons}

\noindent {\bf Procedure A}\\
By using, $N_{\rm pro}( E_{\nu(\bar{\nu})},t,\cos\theta_{\nu(\bar{\nu})}){\rm d}E_{\nu(\bar{\nu})}$,
which is defined in Eq. (\ref{eqn:b2}), 
we define the spectrum for $\cos\theta_{\nu(\bar{\nu})}$  in the following.

\begin{eqnarray}
\lefteqn{      I(\cos\theta_{\nu(\bar{\nu})}){\rm d}(\cos\theta_{\nu(\bar{\nu})}) = } \nonumber \\
&&{\rm d}(\cos\theta_{\nu(\bar{\nu})})\times
\nonumber \\
&&   \times
    \int_{E_{\nu(\bar{\nu}),{\rm min}}}^{E_{\nu(\bar{\nu}),{\rm max}}}
 N_{\rm pro}( E_{\nu(\bar{\nu})},t,\cos\theta_{\nu(\bar{\nu})}){\rm d}E_{\nu(\bar{\nu})}.
 \nonumber \\ 
\label{eqn:c1}
\end{eqnarray}

\noindent By using Eq. (\ref{eqn:c2}) and $\xi$, a sampled uniform random number 
between (0,1), then we can determine $\cos\theta_{\nu(\bar{\nu})}$
from the following equation
    \begin{equation}
      \xi = \int_0^{\cos\theta_{\nu(\bar{\nu})}}P_n(\cos\theta_{\nu(\bar{\nu})})
                          {\rm d}(\cos\theta_{\nu(\bar{\nu})}),
\label{eqn:c2}
    \end{equation}

\noindent where

\begin{eqnarray}
\lefteqn{
P_n(\cos\theta_{\nu(\bar{\nu})}) =  
}
 \nonumber \\
&&
\frac{   I(\cos\theta_{\nu(\bar{\nu})})
} 
{\displaystyle
 \int_0^1 I(\cos\theta_{\nu(\bar{\nu})}){\rm d}(\cos\theta_{\nu(\bar{\nu})})
}.
\nonumber \\
\label{eqn:c3}
    \end{eqnarray}

\noindent {\bf Procedure B}\\
\noindent
For the sampled ${\rm d}(\cos\theta_{\nu(\bar{\nu})})$ in Procedure A,
 we sample 
$E_{\nu(\bar{\nu})}$ from Eq.(\ref{eqn:c4}) by using ${\xi}$, the uniform randum number between (0,1) 

 \begin{equation}
    \displaystyle
       \xi = \int_{E_{\nu(\bar{\nu}),{\rm min}}}^{E_{\nu(\bar{\nu})}} 
                    P_{pro}(E_{\nu(\bar{\nu})},\cos\theta_{\nu(\bar{\nu})}){\rm d}E_{\nu(\bar{\nu})}, 
\label{eqn:c4}
    \end{equation}
where
      \begin{eqnarray}
\lefteqn{
         P_{pro}(E_{\nu(\bar{\nu})},t,\cos\theta_{\nu(\bar{\nu})}){\rm d}E_{\nu(\bar{\nu})} =
} \nonumber \\
&&  
\frac{         N_{\rm pro}( E_{\nu(\bar{\nu})},t,\cos\theta_{\nu(\bar{\nu})}){\rm d}E_{\nu(\bar{\nu})}
}
{\displaystyle         \int_{E_{\nu(\bar{\nu}),{\rm min}}}^{E_{\nu(\bar{\nu}),{\rm max}}} 
         N_{\rm pro}( E_{\nu(\bar{\nu})},t,\cos\theta_{\nu(\bar{\nu})}){\rm d}E_{\nu(\bar{\nu})}
}.
\nonumber\\
\label{eqn:c5}
      \end{eqnarray}

\noindent\\
\noindent {\bf Procedure C}\\
\noindent 
For the sampled $E_{\nu(\bar{\nu})}$ in Procedure B, we sample $E_{\mu(\bar{\mu})}$ from Eqs. (\ref{eqn:a2}) and  (\ref{eqn:a3}). For the sampled  $E_{\nu(\bar{\nu})}$ 
and $E_{\mu(\bar{\mu})}$, we determine $\cos{\theta}_s$, the scattering angle of the muon uniquely from Eq. (\ref{eqn:a1}).\\

\noindent {\bf Procedure D}\\
\noindent
We determine, $\phi$, the azimuthal angle of the scattering lepton from Eq. (\ref{eqn:a5}) by using ${\xi}$, an uniform randum number between (0,1). \\

\noindent {\bf Procedure E}\\
\noindent  
We obtain $\cos{\theta}_{\mu(\bar{\mu})}$ from Eq. (\ref{eqn:a6}).
As the result, we obtain a pair of ($\cos\theta_{\nu(\bar{\nu})}$, 
$\cos{\theta}_{\mu(\bar{\mu})}$) through Procedures A to E. Repeating the 
Procedures A to E, we finally obtain the correlation between the zenith 
angle of the incident neutrino and that of the emitted muon.

\label{}




\newpage
\bibliographystyle{model1a-num-names}
 





\end{document}